\def\nh{{n_H}}

\def\3cm{\rm {cm^{-3}}}
\def\2cm{\rm {cm^{-2}}}
\def\s-1{\rm {s^{-1}}}
\def\etal {et al.}
\def\kms {\hbox{${\rm km\,s}^{-1}$}}
\def\ndv{\hbox{${\rm cm}^{-2}\,{\rm km}^{-1}\,{\rm s} $}}
\def\hcop{\rm {HCO$^+$}}
\def\hcn{\rm {HCN}}
\def\hnc{\rm {HNC}}
\def\cn{\rm {CN}}

\def\nh2{{n(\rm H_2)}}

\documentclass{aa} 
\usepackage{multirow}
\usepackage{graphicx}
\usepackage{epsfig}
\usepackage{txfonts}
\usepackage{psfrag}
\begin{document}
\title{Probing X-ray irradiation in the nucleus of NGC~1068 with observations of high-$J$ lines of dense gas tracers}
\subtitle{}
\author{J.P. P\'{e}rez-Beaupuits\inst{1} \and 
	M.~Spaans\inst{1} \and	
	F.F.S.~van der Tak\inst{2,1} \and	
        S.~Aalto\inst{3} \and 
	S.~Garc\'ia-Burillo\inst{4} \and
        A.~Fuente\inst{4} \and
        A.~Usero\inst{4}}
\offprints{J.P. P\'erez-Beaupuits}
\institute{
 Kapteyn Astronomical Institute, University of Groningen, Landleven 12, 9747 AD Groningen, The Netherlands - 
 \email{jp@astro.rug.nl}
 \and
 SRON Netherlands Institute for Space research, Landleven 12, 9747 AD Groningen, The Netherlands 
 \and
 Onsala Rymdobservatorium, Chalmers Tekniska H\"ogskola, S - 439 92 Onsala, Sweden 
 \and
 Observatorio Astron\'omico Nacional, C/ Alfonso XII 3, 28014, Madrid, Spain 
}
\date{Received  / Accepted  }
\titlerunning{Dense molecular gas in NGC~1068}

%
\abstract
  {Single-dish observations of molecular tracers have suggested that both star formation and an AGN can drive the gas chemistry of the central $\sim$kpc of active galaxies. The irradiation by UV photons from an starburst or by X-rays from an AGN is expected to produce different signatures in molecular chemistry, which existing data on low-$J$ lines cannot distinguish, as they do not trace gas at high temperature and density. Depending on the angular scale of a galaxy, the observed low-$J$ lines can be dominated by the emission coming from the starburst ring rather than from the central region.
}
  {With the incorporation of high-$J$ molecular lines, we aim to constrain the physical conditions of the dense gas in the central region of the Seyfert 2 galaxy NGC~1068 and to determine signatures of the AGN or the starburst contribution.
  }
  {We used the James Clerk Maxwell Telescope to observe the $J$=4--3 transition of HCN, HNC, and HCO$^+$, as well as the CN $N_J=2_{5/2}-1_{3/2}$ and $N_J=3_{5/2}-2_{5/2}$, in NGC~1068. 
  We estimate the excitation conditions of HCN, HNC, and CN, based on the line intensity ratios and radiative transfer models.
  We discuss the results in the context of models of irradiation of the molecular gas by UV light and X-rays.
  }
  {A first-order estimate leads to starburst contribution factors of 0.58 and 0.56 for the CN and HCN $J$=1--0 lines, respectively. 
We find that the bulk emission of HCN, HNC, CN, and the high-$J$ \hcop~emerge from dense gas ($n(\rm H_2)\ge10^5~\3cm$). However, the low-$J$ \hcop~lines (dominating the \hcop~column density) trace less dense ($n(\rm H_2)<10^5~\3cm$) and colder ($T_K\le20$ K) gas, whereas the high-$J$ 
\hcop~emerges from warmer ($>30$ K) gas than the other molecules. 
We also find that the HNC/HCN and CN/HCN line intensity ratios decrease with increasing rotational quantum number $J$.
  }
  {
The \hcop\ $J$=4--3 line intensity, compared with the lower transition lines and with the HCN $J$=4--3 line, support the influence of a local XDR environment.
The estimated $N({\rm CN})/N({\rm HCN})\sim1-4$ column density ratios are indicative of an XDR/AGN environment with a possible contribution of grain-surface chemistry induced by X-rays or shocks.
    }
  

\maketitle

\section{Introduction}

Active galaxies often have high concentrations of molecular gas in
their central ($<$kpc) region, which may be feeding a central active galactic nucleus (AGN) or an starburst (e.g. Sanders \& Mirabel 1996; Gao \& Solomon 1997, 1999).
Molecular gas irradiation by hard X-ray photons ($>$ 1 keV) emitted during the accretion process in the proximity of AGNs, leads to an X-ray dominated region (XDR) (e.g. Maloney \etal\ 1996; Lepp and Dalgarno 1996; Meijerink \& Spaans 2005).
In an starburst region, the formation/destruction of molecules is instead driven by far-UV photons (6 -- 13.6 eV) emitted by O and B stars, which
leave the fingerprint of a photon-dominated region (PDR) (c.f. Hollenbach \& Tilens 1999, and references therein).
An increased X-ray ionization of molecular clouds can enhance the abundance of several molecules (e.g. CN, HNC, HCN and \hcop) with respect to their abundances found in PDRs (e.g. Lepp \& Dalgarno 1996; Meijerink \& Spaans 2005; Meijerink, Spaans \& Israel 2007). Conversely, the chemical state of the molecular gas is a tracer of the dominant energy source of AGNs.

The prototypical Seyfert 2 galaxy NGC~1068 (located at a distance $\sim14.4$ Mpc and with a bolometric luminosity of $\sim10^{11} L_{\sun}$) is a known source where the star formation and/or AGN activity can be debated as the main driving mechanism of its gas chemistry. It hosts a circumnuclear starburst ring of about $2.5-3$ kpc in diameter (Schinnerer \etal\ 2000), which delimits a $\sim2.3$ kpc stellar bar (Scoville \etal\ 1988). 
The detection of strong CO emission coming from the starburst ring indicates that its massive star formation activity is being fed by vast amounts of molecular gas (e.g. Planesas \etal\ 1989, 1991; Helfer \& Blitz 1995; Schinnerer \etal\ 2000). 
However, significant CO emission also arises from a circumnuclear disk (CND) of about 200 pc in diameter (Solomon \& Barret 1991; Schinerer \etal\ 2000).
Earlier observations of high HCN/CO luminosity ratios, measured in the CND of NGC~1068, suggested an abnormal (by Galactic standards) chemistry of molecular gas (Tacconi \etal~\cite{tacconi94}; Sternberg \etal~\cite{sternberg94}). Later observations of HCN, \hcop~and HCO, among other molecules (Usero \etal~\cite{usero04}), suggest that the chemistry in the CND of NGC~1068 corresponds to that of an XDR.
On the other hand, single-dish observations of other high-density tracers (HNC and CN) in the CND of NGC~1068 show that the HNC/HCN and CN/HCN line ratios favor a PDR, rather than an XDR scenario (P\'erez-Beaupuits \etal~\cite{jp07}).
This independent observational evidence for both PDR and XDR scenarios is consistent with a model that considers an AGN and starburst component to be present in NGC~1068 (Spinoglio \etal~\cite{spinoglio05}). However, a later estimate indicates that the age of the most recent episode of star formation, within the central 10--100 pc, is about $250\pm50$ Myr (Davies \etal\ 2007). This result suggests a recent, although no longer active, starburst activity in the CND of NGC~1068.

Nevertheless, it was noted in P\'erez-Beaupuits \etal~(\cite{jp07}) that the result of a favored PDR scenario in the CND of NGC~1068 may be misleading, since the AGN contribution (through XDR effects) is typical of an smaller angular scale than the starburst contribution (PDR). X-rays are known to penetrate deeper than UV photons into a single molecular cloud. But its effects are rather local, since the energy flux of the radiation decreases with radial distance from the source. On the other hand, an starburst region is more spread, and several O/B stars will typically be closer to a molecular cloud than a single X-ray source. Thus the PDR effects will dominate the bulk of the gas. Hence, the smaller angular size of the action zone of an X-ray source (the region around an X-ray source that dominates over UV driven chemical effects) will be affected by beam dilution, and the radiation emerging from the starburst component can dominate the overall emission collected by a single dish beam. 
This is a known drawback of single dish observations of extragalactic sources that, depending on the diameter of the telescope used, complicates the efforts to disentangle the AGN activity from that of the starburst ring, when considering the intensities of the lower transitions of molecular lines.
Given that the starburst contribution is likely dominant at radii $R\sim14''$ in NGC~1068, but not in the CND at $R\sim2''$, the low-$J$ lines observed with single dish telescopes (with beam $>25''$) can be dominated by the contribution from the starburst ring, and if uncorrected, can be misleading.

In this paper we present observations of high-$J$ lines of HCN, HNC, CN and \hcop~on
NGC~1068. Due to the smaller beam size, these lines are less affected by the contribution from 
the starburst ring, than the $J$=1--0 lines. We also present a first order estimate
of the starburst contribution, based on independent observations of the $J$=1--0 lines, and correct 
the intensities of these lines accordingly.
Since HNC is the isomer of HCN, the 
radiation of these molecules is expected to arise from gas with the same physical conditions. 
However, radiative transfer models based on the $\frac{3-2}{1-0}$ line ratios of these
molecules showed that the HNC emissions arise from less dense gas than HCN, when 
considering a kinetic temperature of about 80 K (P\'erez-Beaupuits \etal~\cite{jp07}).
We compare these previous results with those obtained from the 
new $\frac{4-3}{3-2}$ line ratios. The molecular abundances estimated from the radiative 
transfer models are subsequently compared to those obtained with our
PDR and XDR models.

In \S2 we describe the observations. The results (spectral lines and line 
intensities) are presented in \S3. The modelling and analysis 
of the excitation conditions and abundances of HCN, HNC, \hcop~and CN are presented in \S4. 
The conclusions and final remarks of this work are presented in \S5.

\section{Observations and data reduction}

We have used the 16-pixel receiver HARP (Heterodyne Array Receiver Programme) on JCMT\footnote{The JCMT is operated by the Joint Astronomy Center on behalf of the Science and Technology Facilities Council of the United Kingdom, the Netherlands Organization for Scientific Research, and the National Research Council of Canada.}, during December 2006, to observe the $J$=4--3 lines of HNC (363~GHz), HCN (355~GHz) and HCO$^+$ (357~GHz), as well as the CN $N_J=3_{5/2}-2_{5/2}$ (339~GHz), toward the center of the Seyfert galaxy NGC~1068 (r.a.$=02^{\rm h}42^{\rm m}40.7^{\rm s}$ dec.$=-00^{\circ}00'48''$). Beamsizes and efficiencies are shown in Table~\ref{tab:beams}.
We used HARP as a single pixel in staring mode, since it was the only receiver available to cover the high frequencies. The $10^{\rm th}$ pixel was used to point the telescope at the nuclear region of NGC~1068. Detection in the other pixels was negligible and beyond the scope of this work, since they point outside the circumnuclear disk.
Pointing was checked regularly on SiO masers and the RMS was found to be lower than 2$''$. 
Since HARP was not fully commissioned at the time of the observations, we regularly checked 
the calibration of the receiver observing the CO $J$=3--2 line in the reference source CRL168
and we compared it with the representative spectrum obtained with the former receiver B3. 
It was found that the calibration CO $J$=3--2 line observed with HARP was fainter than the reference spectrum, and our data was corrected
according to the variations of the calibration observed during the run. 
We also used the receiver A3 to observe the CN $N_J=2_{5/2}-1_{3/2}$ (226~GHz) line. 
The backend was the ACSIS correlator, and when observing with HARP we used a bandwidth of 1 GHz 
since the 1.8 GHz bandwidth was showing some anomalies in the overlap area and a higher system temperature. 
Unfortunately, this limited our baseline coverage. Nevertheless, we used 2048 spectral channels which gives a resolution of about 0.4 \kms\ in the original spectra. We obtained different numbers of scans (on-source observations) for all the observed lines, with integration times between 5 and 10 minutes each, to allow for calibration observations between the scans. 

   \begin{table}[!t]
      \caption[]{Beam sizes \& efficiencies. }
         \label{tab:beams}
	 \centering
         \begin{tabular}{lcccc}
            \hline\hline
            \noalign{\smallskip}
            Molecule & Transition & $\nu$ & HPBW & $\eta_{\rm mb}$ \\
	             &  $J$/$N_J$ & [GHz] &    [$''$]       &       \\
            \noalign{\smallskip}
            \hline
            \noalign{\smallskip}
            CN & $2_{5/2}-1_{3/2}$  & 226.875 & 21 & 0.67 \\            
            CN & $3_{5/2}-2_{5/2}$  & 339.517 & 15 & 0.63 \\	    
	    HCN & $4-3$ &  354.505 & 14 & 0.63 \\
	    HCO$^+$ & $4-3$ &  356.734 & 14 & 0.63 \\
            HNC & $4-3$ &  362.630 & 14 & 0.63 \\	    
            \noalign{\smallskip}
            \hline
         \end{tabular}
  \end{table}

\begin{figure}[!hp]
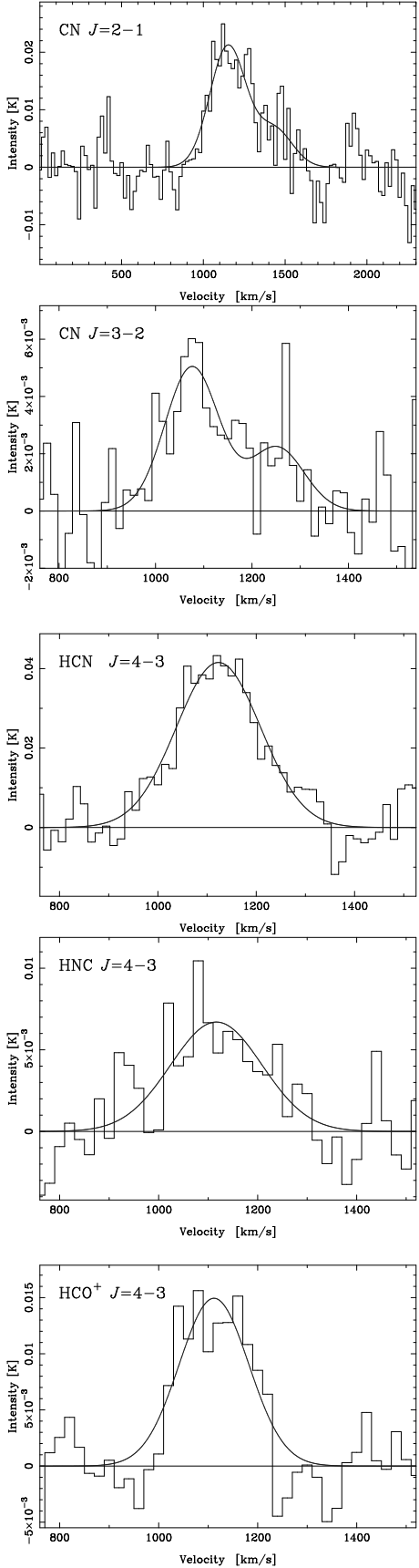


  \hspace*{\fill}\includegraphics[angle=-90,width=6cm]{NGC1068_CN21.ps}\hspace*{\fill}\\
  \hspace*{\fill}\includegraphics[angle=-90,width=6cm]{NGC1068_CN32.ps}\hspace*{\fill}\\
  
  \hspace*{\fill}\includegraphics[angle=-90,width=6cm]{NGC1068_HCN43.ps}\hspace*{\fill}\\
  \hspace*{\fill}\includegraphics[angle=-90,width=6cm]{NGC1068_HNC43.ps}\hspace*{\fill}\\
  
  \hspace*{\fill}\includegraphics[angle=-90,width=6cm]{NGC1068_HCOp43_1gauss.ps}\hspace*{\fill}\\

  \caption{\footnotesize{Molecular line emissions in NGC~1068. 
  The velocity resolution was reduced to 20 \kms for all the lines. 
  The spectra are centered with 
  respect to the heliocentric systemic velocity $\rm v_{sys} = 1137$ \kms. 
  }}
  \label{fig:1068-spectra}
\end{figure}

After a weighted (by system temperature and integration time) average of the scans, a baseline was removed from the average spectra and the resolution was lowered to reduce the level of noise. 
A first-order polynomial was used  in most of the spectra to fit the baselines.
Only for CN $N_J=3_{5/2}-2_{5/2}$ a second order polynomial was needed.
The velocity resolution was reduced to 20 \kms, which represents less than 10\% of the line widths. 
Small variations in the channels chosen to define the baselines in the original averaged spectra, lead to variations of $<6\%$ in the integrated intensities of the final spectra shown in the next section.
The average system temperature was about 300 K for total on-source integration times between 35 minutes and 4.1 hours.
The S/N levels (the peak antenna temperatures, obtained from the gaussian fit, with respect to the r.m.s. of the baselines) ranged between 3 and 7.
The software package XS (written by P. Bergman, Onsala Space Observatory) was used 
to reduce the data and fit the Gaussians.

\section{Results}

The observed molecular line emissions are shown in 
Figure~\ref{fig:1068-spectra} and the line parameters of the gaussian
fits are summarized in Table~\ref{tab:gaussian-fit}.
The spectra are centered with respect to the heliocentric systemic velocity 
$\rm v_{sys} = 1137$ \kms, taken from the NASA/IPAC Extragalactic Database (NED). The velocity-integrated intensities are shown in Table~\ref{tab:intensities}.

   \begin{table}[!tp]
      \caption[]{NGC~1068 line parameters.}
         \label{tab:gaussian-fit}
				\centering
         \begin{tabular}{lcccc}
            \hline\hline
	    \noalign{\smallskip}
            Molecule--$J$ & $V$    & $T_{A}^{*}$ & $\Delta V$ \\
                         & [\kms] &     [mK]    &   [\kms] \\
            \noalign{\smallskip}
            \hline
            \noalign{\smallskip}

	    \multirow{2}{*}{CN $N_J=2_{5/2}-1_{3/2}$} & 1151$\pm$15  & ~~21.4$\pm$~~1.7 & 265$\pm$35\\
		                         & 1444$\pm$42 & ~~~7.1$\pm$~~2.1 & 265$^{~\rm a}$\\

            \noalign{}\\
	    \multirow{2}{*}{CN $N_J=3_{5/2}-2_{5/2}$} & 1076$\pm$14 & ~~~~5.0$\pm$~~0.9  & 132$\pm$34\\
		                         & 1252$\pm$33 & ~~~~2.2$\pm$~~0.8 & ~~~~132$^{~\rm a}$\\
            \noalign{}\\
		HCN $J$=4--3 & 1123$\pm$~~6 & ~~41.5$\pm$~~2.4 & 202$\pm$14\\
            
	    \noalign{}\\
		HNC $J$=4--3 & 1116$\pm$18 & ~~~~6.7$\pm$~~1.1 & 218$\pm$47\\
            
	    \noalign{}\\
		\hcop $J$=4--3 & 1112$\pm$~~8 & ~~~~14.9$\pm$~~1.4 & 166$\pm$19\\
			
            \noalign{\smallskip}
            \hline
         \end{tabular}
\begin{list}{}{}
\item[$^{\mathrm{a}}$] The uncertainty of this parameter was larger than 100\%, if let free. So we set its value to the one found for the main gaussian component.
\end{list}     
   \end{table}
%

\begin{table}[!tp]
      \caption[]{Velocity-integrated intensities.$^{~\rm a}$}
         \label{tab:intensities}	      
     	\centering
         \begin{tabular}{lcc}
            \hline\hline
            \noalign{\smallskip}
            Molecule & Transition & Intensity \\
	             &  $J$/$N_J$ &  [K \kms] \\
            \noalign{\smallskip}
            \hline
            \noalign{\smallskip}
	    CN  & $2_{5/2}-1_{3/2}$ & ~11.7$\pm$~~1.4 \\
            \noalign{\smallskip}	    
	    CN  & $3_{5/2}-2_{5/2}$ & ~~~1.6$\pm$~~0.3 \\
            \noalign{\smallskip}	    
	    HCN & $4-3$ & ~13.9$\pm$~~1.6 \\
            \noalign{\smallskip}	    
	    HNC & $4-3$ & ~~~2.7$\pm$~~0.3 \\
            \noalign{\smallskip}	    
	    HCO$^+$ & $4-3$ & ~~~3.8$\pm$~~0.5 \\
            \noalign{\smallskip}
            \hline
         \end{tabular}
	\begin{list}{}{}
	\footnotesize{
	\item[${\mathrm{a}}$)] Total velocity integrated temperature 
        considering all the components of each spectral line. The values correspond
        to the main-beam brightness intensity, $I_{mb}=\int{T_A^*/\eta_{\rm mb}\delta v}$, in [K \kms].
	}
	\end{list}

\end{table}

The HCN $J$=4--3 spectrum shows the same single peak structure as the lower-$J$
transitions reported by P\'erez-Beaupuits \etal~(\cite{jp07}) and Curran \etal~(\cite{curran00}).
The average line width is about 25\% ($\sim70$ \kms) narrower than the average width observed in the lower 
$J$-transitions.

The double peak structure seen in the HNC $J$=1--0 and $J$=3--2 spectral lines by
P\'erez-Beaupuits \etal~(\cite{jp07}) is not present in the $J$=4--3 line.
This may be due to the smaller beam size and to the lower signal to noise ratio of our data.
The center velocity of the gaussian fit is just about 10~\kms~away from the center 
velocities of the main component in the lower transition lines (within $\sim1\sigma$).
Since this is a single component line, its average width is about 50 \kms~wider than
the line widths of the lower transition lines reported by P\'erez-Beaupuits \etal~(\cite{jp07}), 
but similar to the line width of the HNC $J$=1--0 line reported by H\"uttemeister \etal~(\cite{hutte95}).

The \hcop~spectrum shows a single peak structure, and  appears to be narrower than the $J$=1--0 and 
$J$=3--2 lines reported by Krips \etal~(\cite{krips08}).
Our $J$=4--3 line has an average line width of about 166 km/s, while the lines reported by Krips \etal~have 
average line widths of about 240 km/s. This may be an indication that the $J$=4--3 line traces
a different gas than the lower transitions. But due to the relatively low S/N$\sim6$, deeper observations 
are needed.

Using receiver A3, the two CN spingroups at 226.8746 GHz ($N$ = 2--1, 
\textit{J} = 5/2 -- 3/2, \textit{F} = 7/2 -- 5/2) and 226.6596 GHz 
($N$ = 2--1, \textit{J} = 3/2 -- 1/2, \textit{F} = 5/2 -- 3/2) were detected. 
The central velocity of the second spingroup is about 17 \kms~away from the
expected velocity, which is within the $1\sigma$ error.
In the optically thin limit, the second spingroup is expected to be a factor 
$\le0.56$ weaker than the main spingroup. The factor we obtained from the gaussian 
fit in Figure~\ref{fig:1068-spectra} is $\sim0.33\pm0.1$, which means that
this line is close to optically thin.

Observing CN with HARP at $\sim339.5$ GHz, we also detected a double peak structure.
However, the two spingroups at 339.517 GHz ($N$ = 3--2, \textit{J} = 5/2 -- 5/2, \textit{F} = 7/2 -- 7/2) 
and 339.476 GHz (\textit{F} = 5/2 -- 5/2) are severely blended. The separation between the two peaks is $176\pm36$ \kms, whereas the separation between the farthest spingroup (at 339.447 GHz, $F$ = 3/2 -- 3/2) that we could detect in our bandwidth is expected to be at just about 62 \kms~towards the higher velocities.
So we do not think the double peak structure is due to another spingroup.
Nevertheless, two gaussian components are needed to fit the line profile. Given the noise in the spectrum, the line width of the second component was set to the line width found for the main component.

\section{Discussion}

\subsection{Structure of the CND and limitations of the model}

Millimeter/sub-millimeter and mid-IR high spatial resolution observations of the nuclear region of NGC~1068 support the classical picture of a rotating torus (e.g., Tacconi \etal\ 1994; Schinnerer \etal\ 2000; Galliano \etal\ 2005; Poncelet \etal\ 2007). However, in some previous works (e.g., Schinnerer \etal\ 2000) it was speculated that the classical (large-scale) molecular torus is not always needed to explain the occultation of the nuclear regions. Recent mid-IR observations have shown the existence of two molecular streamers feeding the galactic nucleus of NGC~1068 (Tomono \etal\ 2006; S\'anchez \etal\ 2009). These observations suggest that the occultation of the nuclear region can also be associated with a nuclear concentration of molecular gas and dust that forms an optically thick amorphous clumpy structure (S\'anchez \etal\ 2009). Although it is very likely that this structure encloses smaller infalling clouds, they are not resolved at the 0.075'' (scale size $\sim5$ pc at a distance of 14.4 Mpc) resolution of the Very Large Telescope (VLT).

In the case of millimeter and sub-millimeter interferometer maps of high density tracers (e.g., Tacconi \etal\ 1994; Garc\'ia-Burillo \etal\ 2008), not even the streamers are resolved. This represents a limitation of the current millimeter and sub-millimeter observations in comparison with visual and IR wavelength ranges, with which the morphology and dynamics of the gas can be studied in more detail. This limitation is even more severe in single dish observations for which, depending on the lines observed, the emitting regions (or clouds) can be of a much smaller scale ($\sim1$ order of magnitude) than the beam size of the telescope. In particular, our single dish observations of HCN, HNC, CN and HCO$^+$ are sensitive to the high density ($\sim10^5~\3cm$) clouds that can be embedded even inside the feeding streamers observed by S\'anchez \etal\ (2009). The size of these dense clumps is expected to be from $<1$ pc up to a few pc, embedded in large molecular clouds of the order of tens of parsecs. In the near future, only the Atacama Large Millimeter Array (ALMA) is expected to resolve these smaller clouds at millimeter and sub-millimeter wavelengths with resolutions (at 1mm) between 0.1'' and 0.01'' (scale size $7-0.7$ pc at the distance of NGC~1068).

Ideally, we would like to be able to observe and to model the physical environment in extra-galactic sources, just like it has been done for the Milky Way. Relatively complex radiative transfer codes like, for instance, RATRAN (Hogerheijde \& van der Tak 2000) and $\beta 3D$ (Poelman \& Spaans 2005), can be used to model the internal dynamics, temperature and density structure of individual molecular clouds. However, the detailed structure and signatures of the individual clouds are convolved and smeared out with a large ($>10''$) beam in single dish extra-galactic observations. Therefore, we do not have the information needed to constrain the parameters related to the detailed internal structure of the clouds. What we have is the convolved intensity and line profile of the emission emerging from all the clumps that fall within the telescope beam. Therefore, all modeling efforts are limited to estimate only the bulk of the gas properties. For this purpose, the less complex radiative transfer code RADEX\footnote{http://www.sron.rug.nl/$\sim$vdtak/radex/radex\_manual.pdf} (Van der Tak \etal~\cite{vdtak07}) is a sensitive model for estimating these average physical conditions of the high density gas from single dish observations towards extra-galactic sources. However, due to the lack of spatial resolution in our data, and the limitations of our model, the following discussion and the conclusions derived from them should be taken as qualitative orders of magnitude. Higher spatial resolution, such as the resolutions that ALMA will provide, are necessary to reach definitive conclusions.

\subsection{Line ratios and starburst contribution}

The observed line ratios of molecular transitions have been largely used as a tool for estimating the physical conditions of the molecular gas. However, the conversion from $T_{\rm mb}$ to $T_R$ is uncertain through the source size of the emitting region and, hence, the derived physical conditions ($T_K$, $n(\rm H_2)$ and $N$) depend strongly on this conversion.
Special care regarding this issue has been taken in previous work (e.g. Tacconi \etal~1994, Usero \etal~2004, P\'erez-Beaupuits \etal~2007, Krips \etal\ 2008).

The high resolution map of the CO $J$=1--0 line in NGC~1068 reported by Schinnerer \etal~(2000) shows that the morphology of this galaxy consists mostly of a nuclear region (circumnuclear disk) and spiral arms (starburst ring). 
Since our aim is to use the HCN, CN, HNC and \hcop\ data as a diagnostic tool for AGN-like environments, we are interested only in the emission emerging from the CND of NGC~1068. However, the starburst ring could potentially be contributing to the observed emission and, hence, influence the results of the radiative transfer models, as noted by P\'erez-Beaupuits \etal~(2007). This is particularly relevant for the $J$=1--0 lines of all the molecules because of the larger beam size. Therefore, the line intensity ratios, involving the $J$=1--0 transitions, could be misleading. 

A priori, the most straightforward way to estimate the contribution from the starburst ring would be to use interferometer data, for instance, from the HCN $J$=1--0 map reported by Tacconi \etal~(1994). Although this is possible, using interferometer data is somewhat delicate due to the missing flux. Even though the zero spacing problem can be resolved through spectra observed with a single-dish telescope in the direction of the interferometer phase center, only part of the short spacing information can be recovered. Besides, high quality interferometer and single-dish data are required in order to minimize the artifacts in the final combined data. Because the uncertainties in this technique are higher, and because high quality data are not available for all the lines needed, we derive below a first-order estimate of the starburst contribution, based on single-dish data only. We use the available $J$=1--0 lines of HCN, HNC and CN observed with different telescopes (SEST, OSO, IRAM 30m). 

The total flux or intensity $F_{\rm TOTAL}$, measured with a beam of FWHM=$\theta_{\rm mb}$ that includes emission from the starburst ring ($F_{\rm SB}$), of average radius $r\sim14''$, and from the CND ($F_{\rm CND}$, with an smaller radius of about 1''), can be decomposed as follows:

\begin{equation}
 F_{\rm TOTAL}=F_{\rm CND}*w_{\rm CND}+F_{\rm SB}*w_{\rm SB} ,
\end{equation}

\noindent
where $w_{\rm CND}$ and $w_{\rm SB}$ are weighting functions that can be estimated as $w_{\rm CND}\sim1$, because the size of the CND is small compared to the beams of any of the available single dish data, and

\begin{equation}
 w_{\rm SB}\sim exp\left[ -4*ln(2)*\left( \frac{14''}{\theta_{\rm mb}}\right) ^2\right] .
\end{equation}

In eq. (2) we assume that the contribution comes \textit{only} from an axisymmetric ring-like distribution of radius $r=14''$, neglecting the contribution from the bar or any other feature at intermediate radii. Since we have at least two \textit{independent} observations of the $J$=1--0 line, with different telescopes of beams $\theta_{\rm mb1}$ and $\theta_{\rm mb2}$, we can have a first-order estimate of the starburst contribution as:

\begin{equation}
 F_{\rm SB}\sim \frac{F_{\rm TOTAL1}-F_{\rm TOTAL2}}{w_{\rm SB1}-w_{\rm SB2}} .
\end{equation}

Hence, from eq. (1), the starburst contribution factor $f_{\rm SB}$ of a line observed with any telescope can be estimated as:

\begin{equation}
 f_{\rm SB}\sim \frac{F_{\rm SB}*w_{\rm SB}}{F_{\rm TOTAL}} .
\end{equation}

So we can estimate the intensity that emerges from the CND from the observed total intensity as $F_{\rm CND}\sim F_{\rm TOTAL}*(1-f_{\rm SB})$. 
This means that, the larger $f_{\rm SB}$, the larger will be the $\frac{3-2}{1-0}$ ratios.
Note that, from equation (2), $f_{\rm SB}$ depends on the beam size $\theta_{\rm mb}$ of the telescope used for the observations.
Using the peak intensities from P\'erez-Beaupuits \etal~(2007) and from available IRAM 30m data for the same CN, HCN and HNC $J$=1--0 lines (Usero \etal\ 2009) and the corresponding telescope beams, we estimate the $f_{\rm SB}$ factors as $0.58\pm0.25$ for CN (SEST, $\theta_{\rm mb}=45''$), $0.56\pm0.32$ for HCN (OSO, $\theta_{\rm mb}=44''$) and $0.99\pm0.50$, for HNC (SEST, $\theta_{\rm mb}=55''$). 

The similar (within $\sim4$\%) starburst contribution factors found for CN and HCN, as observed with SEST and OSO, respectively, are as expected since the respective beam sizes are also similar (within $\sim2$\%). Instead, the high contribution factor estimated for HNC would indicate that only 1\% of the observed emission emerges from the nuclear region, in contrast to the 42\% and 44\% found for CN and HCN. If this was real, we would expect to see a pronounced double peak structure, with the proper velocity centers, in the HNC $J$=1--0 line profile. Since this is not the case, we think that the substantial difference between the starburst contribution factors of HNC and the other molecules might be due to some problems (e.g., pointing or calibration errors) in the spectral line reported in P\'erez-Beaupuits \etal~(2007), which was noted to be different than the one previously reported by H\"uttemeister \etal~(1995). Thus, we use the HNC $J$=1--0 line from H\"uttemeister \etal\ for the following analysis. However, this line, as well as the low-$J$ \hcop~lines reported by Krips \etal~(2008), was observed with the IRAM 30m telescope only. Therefore, we cannot estimate the corresponding starburst contribution factor, due to the lack of independent observations. Hence, for HNC and \hcop\ $J$=1--0 lines we adopt the value $f_{\rm SB}=0.45\pm0.31$ estimated from the HCN observations done with the 30m telescope ($\theta_{\rm mb}=28''$).
The starburst contribution factors for the higher-$J$ transitions are expected to be negligible because of the smaller beams at higher frequencies, so we can assume these to be zero.

\begin{table}[!t]
      \caption[]{Line intensity ratios between the peak antenna temperatures, assuming a source size of 1.5''.}
         \label{tab:line-ratios}
	\centering
         \begin{tabular}{lcc}
            \hline\hline
            \noalign{\smallskip}
            Molecule & Transitions & Ratio$^{~\mathrm{a}}$ \\
            \noalign{\smallskip}
            \hline
            \noalign{\smallskip}
	    CN  & $\frac{2_{5/2}-1_{3/2}}{1_{3/2}-0_{1/2}}$ & 0.61$\pm$0.39 \\
            \noalign{\smallskip}
	    CN  & $\frac{3_{5/2}-2_{5/2}}{2_{5/2}-1_{3/2}}$ & 0.13$\pm$0.04 \\
            \noalign{\smallskip}	    	    
	    HCN & $\frac{3-2}{1-0}$ & 1.01$\pm$0.53 \\
            \noalign{\smallskip}	    
	    HCN & $\frac{4-3}{3-2}$ & 0.49$\pm$0.14 \\
            \noalign{\smallskip}	    
	    HNC & $\frac{3-2}{1-0}$ & 0.38$\pm$0.18 \\
	    \noalign{\smallskip}
	    HNC & $\frac{4-3}{3-2}$ & 0.30$\pm$0.10 \\
	    \noalign{\smallskip}
	    HCO$^+$ & $\frac{3-2}{1-0}$ & ~~~0.12$\pm$0.04$^{~\mathrm{b}}$ \\
            \noalign{\smallskip}	    	    
	    HCO$^+$ & $\frac{4-3}{3-2}$ & 1.37$\pm$0.46 \\
            \noalign{\smallskip}	    	    
            \hline
         \end{tabular}
\begin{list}{}{}
\item[${\mathrm{a}}$)] See P\'erez-Beaupuits \etal~(\cite{jp07}) for the $J$=1--0 and $J$=3--2 transitions. The low-$J$ lines of \hcop\ were obtained from Krips \etal~(2008).
\item[${\mathrm{b}}$)] This ratio slightly differs from that found by Krips \etal\ (2008) because we use the ratio between the peak main beam temperatures and we corrected the \hcop~$J$=1--0 line for the starburst contribution factor.
\end{list}
\end{table}
%

\begin{table*}[!t]
      \caption[]{Physical conditions estimated from line ratios$^{\mathrm{a}}$ and line intensities.}
         \label{tab:xc}
	\centering
         \begin{tabular}{lcccc}
            \hline\hline
            \noalign{\smallskip}
            Molecule/ratio & $\nh2$ & $T_K$ & $N/\Delta\upsilon^{\mathrm{b}}$ & $\tau^{\mathrm{b}}$ \\
                           & [$\3cm$] & [K] & [$\ndv$] & \\
            \noalign{\smallskip}
            \hline
            \noalign{\smallskip}

            HCN & $10^4-10^7$ & $>20$ & $10^{13.2}-10^{15.2}$ & $10^{-3.0}<\tau_{1-0}<10^{2.4}$ \\
                &             &       &                       & $10^{0.2}<\tau_{3-2}<10^{2.8}$ \\
                &             &       &                       & $10^{0.2}<\tau_{4-3}<10^{2.8}$ \\
            \noalign{\smallskip}
            \hline
            \noalign{\smallskip}
            HNC & $10^4-10^7$ & $10-90$ & $10^{12.8}-10^{15.0}$ & $10^{-1.0}<\tau_{1-0}<10^{2.6}$ \\
                &             &       &                         & $1<\tau_{3-2}<10^{2.8}$ \\
                &             &       &                         & $10^{-0.6}<\tau_{4-3}<10^{2.4}$ \\
            \noalign{\smallskip}
            \hline
            \noalign{\smallskip}
            CN & $10^{5.2}-10^7$ & $10-20$ & $10^{14}-10^{15.2}$ & $10^{-2.0}<\tau_{1_{3/2}-0_{1/2}}<10^{2.4}$ \\
                &             &       &                          & $1<\tau_{2_{5/2}-1_{3/2}}<10^{2.8}$ \\
                &             &       &                          & $10^{-0.4}<\tau_{3_{5/2}-2_{5/2}}<10^{1.4}$ \\
            \noalign{\smallskip}
            \hline
            \noalign{\smallskip}

            \hcop $\frac{4-3}{3-2}$ & $10^{5.9}-10^7$ & $>30$ & $10^{11.9}-10^{12.2}$ & $10^{-4.0}<\tau_{3-2}<10^{-1.2}$ \\
                &             &       &                         & $10^{-2.2}<\tau_{4-3}<10^{-1.2}$ \\

            \noalign{\smallskip}	    	    
            \hline
         \end{tabular}
\begin{list}{}{}
\item[${\mathrm{a}}$)] Results from the single-phase model with a source size $\theta_{\rm S}=1.5''$.
\item[${\mathrm{b}}$)] In order to get the actual column density and optical depth, the values reported here must be multiplied by a pertinent line width, i.e. thermal plus turbulent width (FWHM) of the lines. According to Schinnerer \etal~(2000), the velocity dispersion at the nuclear region of NGC~1068 is about 23~\kms.
\end{list}
\end{table*}

After correcting the $J$=1--0 lines for the starburst contribution, we correct the peak intensities for beam dilution as 
$T_R=\frac{ T_A^* }{ \eta_{\rm mb} }\times\frac{ \theta_{\rm S}^2+\theta_{\rm mb}^2 }{ \theta_{\rm S}^2}$.
The ratios between the line intensities are shown in Table~\ref{tab:line-ratios}.
We use the ratios between the peak intensities of the \textit{main components} in the Gaussian fit
shown in Table~\ref{tab:gaussian-fit}. The peak intensity corresponds to the emission emerging from regions with velocities close to the systemic velocity of NGC~1068, and the ratios obtained from these values should be similar to those obtained from the velocity-integrated intensities, if the line widths of the corresponding components of the different transitions are also similar.
The peak intensities of the low-$J$ transitions of HCN and CN were taken from P\'erez-Beaupuits \etal~(\cite{jp07}). 
The peak main beam temperatures of the lower transitions of \hcop\ were provided by M. Krips (private communication; Krips 
\etal~\cite{krips08}).

The line ratios were computed considering uncertainties of 10\% in the reported beam efficiencies $\eta_{mb}$, 5\% in the main beam $\theta_{mb}$, and 10\% in the estimated source size $\theta_S$. As described in the next section, $\theta_S$ was assumed to be 1.5'' and the same for all the lines. Note that the high uncertainties given in Table~\ref{tab:line-ratios} should be considered lower limits, since comparison of data observed with different telescopes may introduce additional uncertainties due to the presence of different unknown systematics between the telescopes.

\subsection{Modeling the excitation conditions}

We explore a wide range of possible excitation conditions 
that can lead to the observed line intensities and line ratios, in a similar way as described in
P\'erez-Beaupuits \etal~(\cite{jp07}). 
In the radiative transfer models, the kinetic temperature $T_K$ and the hydrogen density $n(\rm H_2)$ can be constrained from the \textit{line ratios} (Table~\ref{tab:line-ratios}), whereas the column density (per line width) is mostly constrained by the \textit{line intensities}. We used the intensity of the $J$=3--2 line of HCN, HNC, and \hcop, and the $N_J=2_{5/2}-1_{3/2}$ line of CN, since these are common in the ratios between the low-$J$ and high-$J$ lines. The intensity of these lines was corrected by beam dilution as described below.
We use the radiative transfer code RADEX to build a cube of kinetic temperature $T_K$, number density $n(\rm H_2)$ and column density per line width $N/\Delta V$ for each molecule. 

For most of the molecules we explored collisions only with $\rm H_2$, since this is considered to be the most important astrophysical process. Other collision partners can be H and He. Although their collision cross sections are comparable, $\rm H_2$ is about 5 times more abundant than He, and H is at least one order of magnitude less abundant than $\rm H_2$ in the dense cores of molecular clouds (e.g., Meijerink \& Spaans 2005). Hence, the effects of collisions with these other partners can be neglected. Nevertheless, in contrast with HCN, HNC and ${\rm HCO^+}$, the open shell molecular structure of the CN radical makes it more interactive with electrons. Therefore, we also explore the CN--$e^-$ collision as an additional process that could aid the excitation of CN, since the electron abundance can be enhanced by a high ionization degree in PDRs and XDRs.

The volume density explored ranges between $10^4~\3cm$ and $10^7~\3cm$, the kinetic temperature varies from 4 K to 200 K, and the column density per line width lies between $10^{10}$ \ndv~and $10^{18}$ \ndv. 
In order to get the actual column density and optical depth, the values that will be reported here must be multiplied by a line width. According to Schinnerer \etal~(2000), the velocity dispersion at the nuclear region of NGC~1068 is about 23~\kms. So, we could expect local velocity dispersions (line widths) ranging between 3~\kms and 10~\kms.

The original RADEX code was modified in order to include dust background emission as a diluted blackbody radiation field, in the same way as done by Poelman \& Spaans (2005). We consider the emission component at 34 K described in Spinoglio \etal\ (2005), scaled by the continuum optical depth $\tau_c(\nu)$ at a particular frequency. Since in our models we consider up to ten transitions, we estimated $\tau_c(\nu)$ as increasing linearly with the frequency of the transition (from Fig.4c in Spinoglio \etal\ 2005). We modeled the total background radiation as a composite between the Cosmic Background Radiation (CMB), as a blackbody function at 2.73 K, and the diluted dust radiation, which was estimated as $\tau_c\times B(T_{dust})$, where $B(T_{dust})$ is the Planck function at $T_{dust}=34$ K and $\tau_c(\nu)=\tau_{100\mu{\rm m}}(100\mu{\rm m}/\lambda)$. 
From Hollenbach \etal\ (1991) we adopted the value $\tau_{100\mu {\rm m}}=10^{-3}$, hence the continuum optical depth becomes $\tau_c(\nu)=10^{-5}{\rm [cm]}\nu/c$, where $\nu$ is the rest frequency of the transition in [Hz], and $c$ is the speed of light in [cm/s].

The inclusion of the diluted continuum emission does not affect the results found for \hcop\, but decreases (from $10^{4.7} \3cm$ to $10^{5.2} \3cm$) the minimum density at which solutions are found for CN, and it increases (by $\sim10-20$ K) the minimum temperatures found for HCN and HNC, at the lowest densities, in comparison with a model that considers only the CMB as background radiation. The lower limits of the column densities per line width increases as well.

The physical conditions of HCN, \hcop, and HNC were modeled using the collisional data available in the LAMDA\footnote{http://www.strw.leidenuniv.nl/$\sim$moldata/} database (Sch\"oier \etal~\cite{schoier05}). In the case of CN, we used the collision rates estimated by Fuente \etal~(1995). We do not consider hyperfine splitting of CN, HNC and HCN, since they are not resolved in extragalactic systems. 

In P\'erez-Beaupuits \etal~(\cite{jp07}) the excitation conditions of HCN and HNC overlapped, assuming that their emission emerges from the same region. Since we now have more transitions, we can constrain the temperature and density for each molecule separately, and then see if they agree or not.

Once the starburst contribution to the low-$J$ lines is removed, the bulk of the emission for all the dense gas tracers mostly arises from the CND. There should not be a significant difference in the sizes of the CND as seen in HCN, \hcop, HNC and CN, if all the lines can be modeled by a single phase. So we assume the same source size for all the $J$ lines to correct them for beam dilution.
From the interferometer maps of CN, HCN and \hcop~available in the literature for NGC~1068 (e.g. Tacconi \etal~1994, Kohno \etal~2001, Garc\'ia-Burillo \etal~2008), we estimate that the source size (FWHM) of these molecules ranges between about 1 and 2 arcseconds. 

We can model the physical conditions for most of the line ratios of most of the molecules, considering an average source size $\theta_{\rm S}=1.5''$. A summary of the estimated physical conditions is provided in Table~\ref{tab:xc}, and a summary of the main results is presented in the next section. Details of the modeling and analysis of the uncertainties are given in the Appendix.

\subsection{Physical environment and abundance ratios}

All the solutions for HCN are found for temperatures higher than 20 K, with a degeneracy between the kinetic
temperature and molecular hydrogen density.
The temperatures at which solutions can be found for HNC are limited up to about 90 K.

The solutions found for both CN $\frac{2-1}{1-0}$ and $\frac{3-2}{2-1}$ line ratios overlap in an small region, with a narrow temperature range (between 10 K and 20 K). However, if we use an smaller source size the temperature range can go from 20 K up to 100 K. The analysis of the uncertainties in CN is presented in Appendix B.1.
We found that the effect of electrons as a secondary collision partner of CN, is not important in a PDR environment due to the relatively low electron abundance ($\sim10^{-4}$) with respect to H$_2$ and because the abundance of CN is very low where $n(e^-)$ is considerably higher (at the outer layers of a molecular cloud, close to the ionization front). In an XDR environment, small effects can be expected from collisions with electrons, mainly for the high-$J$ lines. 

Using $\theta_{\rm S}=1.5''$ our model is not able to reproduce the HCO$^+$ $\frac{3-2}{1-0}$ line ratio and the HCO$^+$ $J$=3--2 line intensity simultaneously. But it is able to reproduce the HCO$^+$ $\frac{4-3}{3-2}$. 
However, our model can reproduce the observed HCO$^+$ $\frac{3-2}{1-0}$ ratio, and HCO$^+$ $J$=3--2 intensity, if we assume either a larger source size, or a larger starburst contribution factor. Detailed analysis of these alternatives is described in Appendix B.2.

We estimate an $N(\hnc)/\textit{N}(\hcn)$ column density (abundance) ratio lower than unity in the CND of NGC~1068. This result cannot be interpreted as a signature of a pure PDR or XDR environment because the HNC abundance can be decreased due to high temperatures produced inside a molecular cloud by mechanisms like turbulence and shocks (Loenen \etal~\cite{loenen08}). The high SiO abundance observed in the CND of NGC~1068 can be a direct evidence of the possible contributions from mechanical heating (shocks) and dust grain chemistry induced by X-rays (Garc\'ia-Burillo \etal\ 2008). Hence, a more elaborate PDR/XDR model that includes both mechanical heating and grain surface chemistry is needed for the interpretation of our results.

The mean $N({\rm CN})/N({\rm HCN})$ column density ratios (0.7--0.9) estimated with $\theta_{S}=1.5''$ can be easily found in an XDR environment, but the predominance of this component cannot be concluded from the $N({\rm CN})/N({\rm HCN})$ ratio only, since ratios $\sim1.0$ are also expected in a PDR component (Lepp \& Dalgarno 1996, Meijerink \etal~\cite{meijerink07}). The mean column density ratios $2\lesssim N({\rm CN})/N({\rm HCN})\lesssim 4$ estimated with $\theta_{S}=1''$, are tentatively more consistent with an XDR/AGN environments (Lepp \& Dalgarno 1996, Meijerink \etal~\cite{meijerink07}). However, these and higher [CN]/[HCN] abundance ratios have also been found in PDR/starburst scenarios (e.g. Fuente \etal\ 2005). The fact that we find a relatively low $N({\rm CN})/N({\rm HCN})$ column density ratio, with respect to what would be expected in a pure XDR scenario, could be explained by an overabundance of HCN due to the grain-surface chemistry suggested by Garc\'ia-Burillo \etal\ (2008).

Our model indicates that the emission from the high-$J$ HCO$^+$ lines emerge from gas that does not co-exist with HCN, HNC and CN, in the nuclear region of NGC~1068. The warmer and denser gas traced by the high-$J$ lines is only an small fraction (0.5\% -- 10\%) of the total HCO$^+$ gas. Most of it is confined to the lower transitions. Details of the analysis and interpretation of the abundance ratios are presented in Appendix C.

\section{Conclusions}

We have used HARP on JCMT to observe the $J$=4--3 line of 
HCN, HNC, \hcop~, and the $N_J=2_{5/2}-1_{3/2}$ and $N_J=3_{5/2}-2_{5/2}$ transitions of CN, in the nuclear region of NGC~1068.
We estimated the excitation conditions and abundance ratios of these molecules in the CND from radiative transfer models, assuming a single phase gas for all the molecules and lines. We compared column density ratios with predictions from PDR/starburst and XDR/AGN models.

We deduced a first-order estimate of the starburst contribution to the $J$=1--0 lines, out of independent observations with different telescopes. The starburst contribution of the higher-$J$ lines should be negligible because of the smaller beam sizes at the higher observing frequencies.

We first expected, and assumed, that the source sizes, seen by the different molecules and transitions, should not be significantly different. We estimated the source sizes for HCN, CN and \hcop~from the different interferometer maps available in the literature, and used a common value of 1.5'' for all the observed lines. The results of our models showed that the HCN, HNC and partly CN, can co-exist in gas with similar kinetic temperature (10 -- 20 K) and hydrogen density ($10^{5.3-5.7}~\3cm$).

However, the analysis of the uncertainties in source size, as well as in the starburst contribution factor, shows that the \hcop~molecule seems to trace a totally different gas phase, both with respect to the other molecules and its own different transitions. The \hcop~$\frac{4-3}{3-2}$ line ratio indicates that the emission of these lines emerges from gas with densities larger than $10^{5.9}~\3cm$ and temperatures higher than 30 K. Instead, the \hcop~$\frac{3-2}{1-0}$ line ratio would traces less dense ($n(\rm H_2)\le10^5~\3cm$) and colder ($T_K\le20$ K) gas, when using models with either an smaller source size or a higher starburst contribution factor.

The physical conditions estimated from the line ratios are based on the assumption of a single-phase model with the same size of the emitting region. However, a single-phase scenario clearly does not hold for the \hcop~molecule. The lack of an independent measurement of the \hcop~$J$=1--0 line, and the assumption of the same starburst contamination factor for this line as the one found for HCN from the IRAM 30m data, introduces an additional uncertainty. Nevertheless, the fact that the high-$J$ lines of \hcop~clearly trace higher densities and temperatures, with lower \hcop~column densities, indicates that this gas is less extended throughout the CND, in terms of volume and maybe in surface as well. 

The line intensity ratios of CN/HCN and HNC/HCN decrease with increasing rotational quantum number $J$. The \hcop/HCN intensity ratios, instead, do not follow this trend. The \hcop\ $J$=4--3 line intensity, compared with the lower transition lines and with the HCN $J$=4--3 line, support the influence of a local XDR environment.

The $N({\rm CN})/N({\rm HCN})\sim1-4$ column density ratios estimated with the single-phase model are indicative of an XDR/AGN environment with a possible contribution of grain-surface chemistry induced by X-rays or shocks.

The single-phase model, with a common source size seen by all the molecules and transitions, works well for HCN and HNC. However, \hcop~shows clear evidence that different lines can trace completely different gas phases. Besides, assuming an average source size for all the molecules and lines can be a good first approximation, but has the clear drawback that estimates derive mostly from the $J$=1--0 interferometer maps, which do not trace very warm and dense gas, and are sensitive to beam smearing effects. High resolution maps of the higher-$J$ lines may show slight differences in source sizes that can be considered negligible, at a first glance. But, as was shown with CN and \hcop, even an error of a fraction of an arcsecond in the estimated source size can make a big difference in the physical conditions derived from the radiative transfer models. High resolution maps of the different transitions used in the models would aid to reduce the uncertainties in the models. In addition, a more elaborated PDR/XDR model that includes mechanical heating and grain-surface chemistry, will be necessary to properly account for the different contributing factors.

The spatial extent and morphology of the emitting regions of different $J$ lines, as well as the detailed dynamics and physical environment of the high density tracers, can be tested with the higher spatial resolution maps that ALMA will provide in the future.

\acknowledgements{Many thanks go to Kalle Torstensson for doing the
observations, and to the JCMT staff for their support. We are 
grateful to J.H. Black for useful comments and M. Krips for providing 
\hcop\ data. We are grateful to the referee for the careful 
reading of the manuscript and constructive comments and suggestions.
Molecular Databases that have been helpful include 
the NASA/JPL, LAMDA and NIST. This work has benefitted from research 
funding by the European Community's sixth Framework Programme under 
RadioNet R113CT 2003 5058187.}

\Online

\begin{appendix}
\normalsize

\section{Modeling the physical conditions of high density tracers}

\subsection{The physical environment of \hcn}

\begin{figure}[!tp]
  \centering
  \includegraphics[width=7cm]{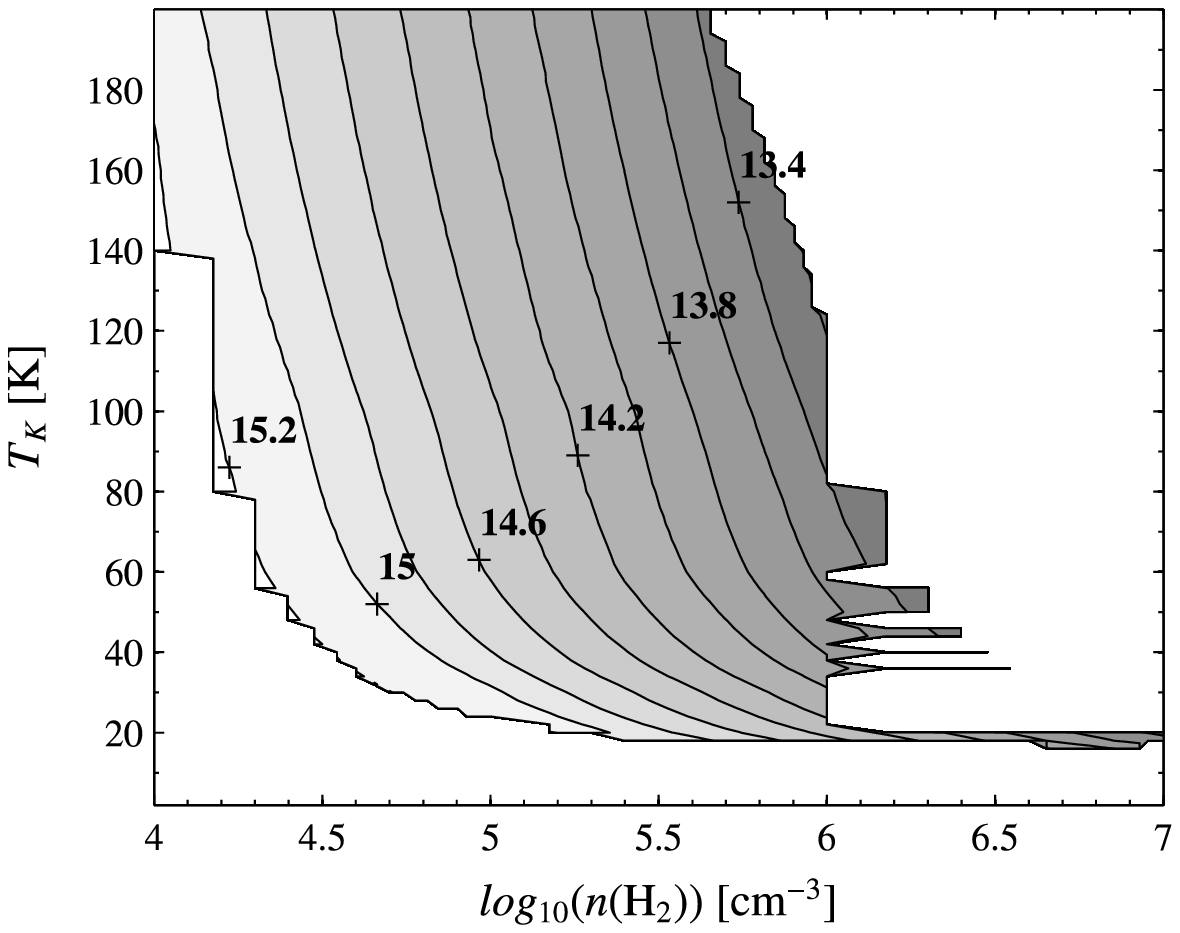}\\
  \includegraphics[width=7cm]{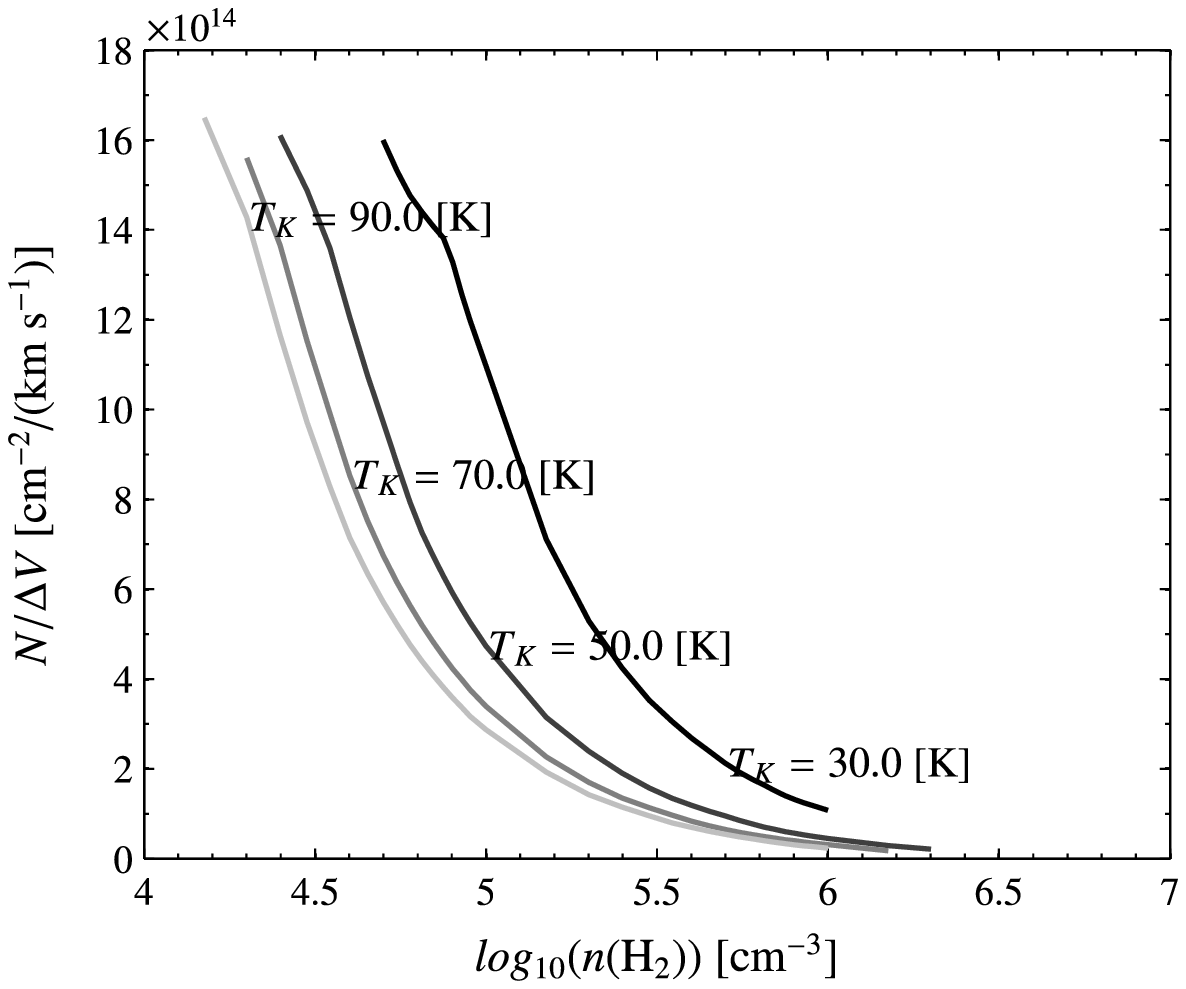}%
  
  \caption{\textit{Top} - Excitation conditions modeled for the $\frac{3-2}{1-0}$ 
  and $\frac{4-3}{3-2}$ line ratios of HCN. The contour lines correspond
  to the mean column density of HCN per line width ($N(\rm{HCN})/\Delta\upsilon~\ndv$), 
  in the region where the estimated line ratios and the $J$=3--2 line intensity reproduce,
  within $1\sigma$, the observed values. \textit{Bottom} - Transversal cuts of the column density per line width at different temperatures modeled above. The column densities are of the order of $10^{14}~\ndv$, in linear scale.}
  \label{fig:HCN-xcmaps}
\end{figure}

The \textit{top panel} of Figure~\ref{fig:HCN-xcmaps} shows the mean column density per line width  
$N(\rm{HCN})/\Delta\upsilon~\ndv$ (the mean between the maximum and minimum $N/\Delta\upsilon$ that yield valid solutions), for all the explored densities $n(\rm H_2)$ and temperatures 
$T_K$ that can reproduce, within $1\sigma$, the observed HCN $\frac{3-2}{1-0}$ and HCN $\frac{4-3}{3-2}$ 
line ratios and the intensity of the HCN $J$=3--2 line. The \textit{bottom panel} of 
Figure~\ref{fig:HCN-xcmaps} shows how $N(\rm{HCN})/\Delta\upsilon~\ndv$ changes with $\nh2$ at
different temperatures. These curves are easier to compare with the output of Large Velocity Gradient (LVG) models 
(e.g. Goldreich \& Scoville 1976) commonly found in the literature. To make these curves more clear
we do not show the error bars (or uncertainties) of $N(\rm{HCN})/\Delta\upsilon$ for each temperature in the plot. 
But they range between 25.0\% -- 48.8\% at 30 K, 24.2\% -- 40.6\% at 50 K, 
18.4\% -- 38.0\% at 70 K, and 21.2\% -- 36.4\% at 90 K. Note that the uncertainties decrease towards the higher $\nh2$ densities.

The lower and upper limits of the HCN column density per line width $N(\rm{HCN})/\Delta\upsilon~\ndv$ are 
$10^{13.2}~\ndv$, at the highest density of about $10^6~\3cm$, and $10^{15.4}~\ndv$ at the lowest density 
explored of $10^4~\3cm$, respectively. There is also a narrower temperature region (around 20 K) with
densities higher than $10^6~\3cm$, where solutions for the observed ratios and intensities are also possible.
The optical depths $\tau$ of each line are summarized in Table~\ref{tab:xc}.

All the solutions are found for temperatures higher than 20 K, with a clear degeneracy between the kinetic
temperature and molecular hydrogen density. That is, a given column density can be obtained with either high $T_K$ and low $\nh2$,
or low $T_K$ and high $\nh2$.

\subsection{The physical environment of \hnc}

In the case of the double peak structure observed in the lower transition lines of HNC,
only the main peak (at velocity $\sim1073\pm13$ \kms) was considered in the analysis, since
this is the component that is closer to the central velocity observed in the HNC $J$=4--3 line.

\begin{figure}[!tp]
  \centering
  \includegraphics[width=7cm]{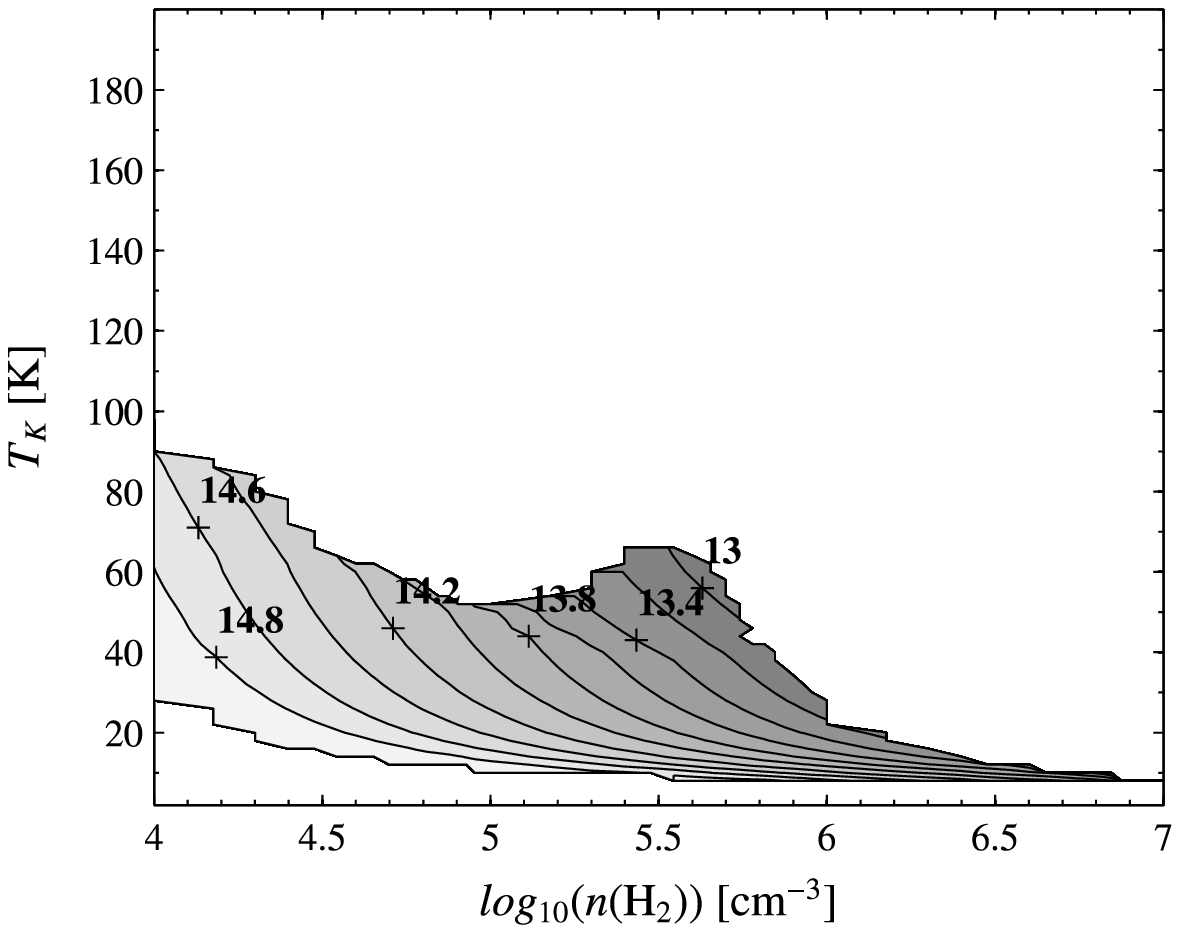}\\
  \includegraphics[width=7cm]{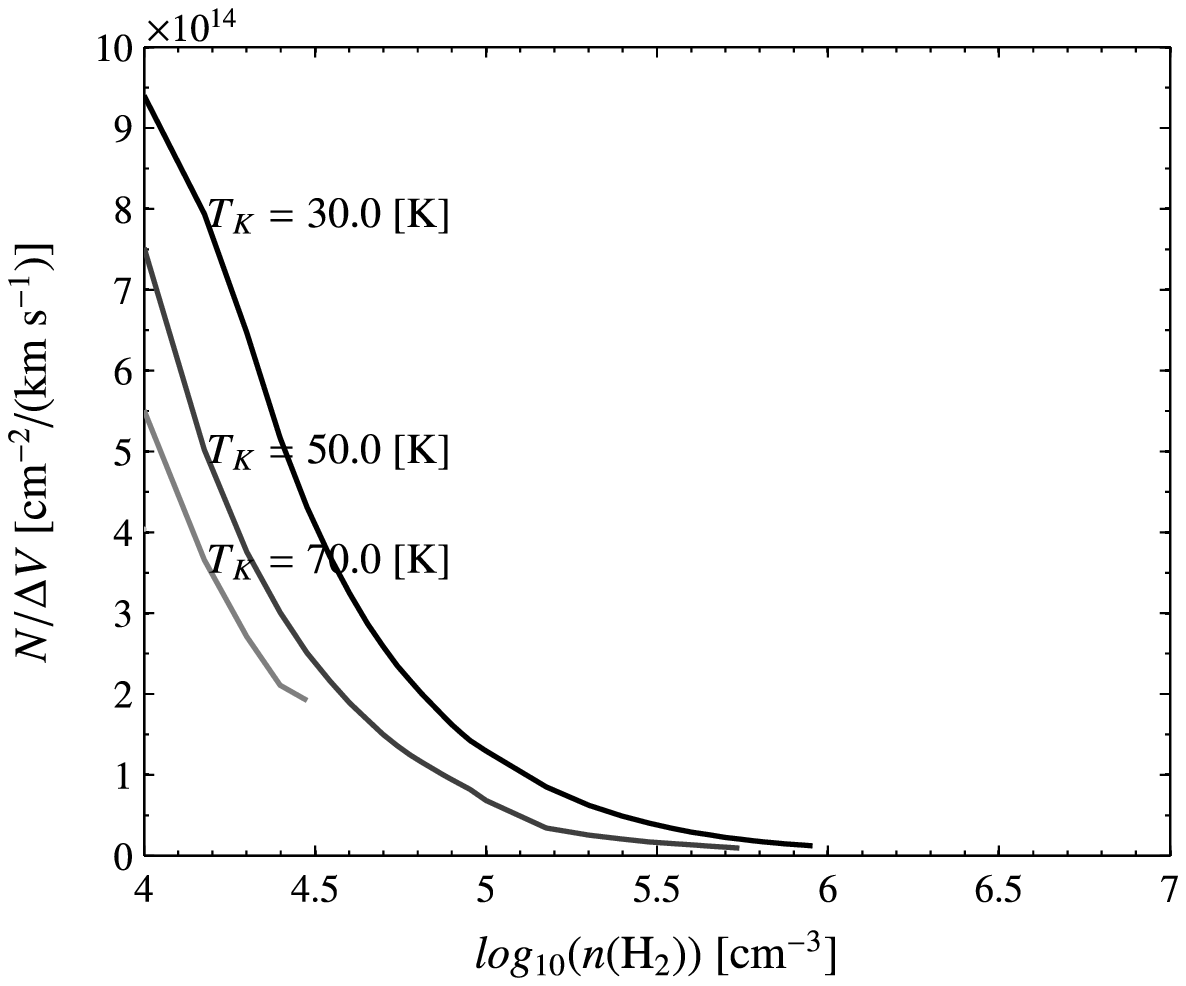}
  
  \caption{\textit{Top} - Excitation conditions modeled for the $\frac{3-2}{1-0}$ 
  and $\frac{4-3}{3-2}$ line ratios of HNC. The contour lines correspond
  to the mean column density of HNC per line width ($N(\rm{HNC})/\Delta\upsilon~\ndv$), 
  in the region where the estimated line ratios and the $J$=3--2 line intensity reproduce,
  within $1\sigma$, the observed values. \textit{Bottom} - Transversal cuts of the column density per line width at different temperatures modeled above. The column densities are of the order of $10^{14}~\ndv$, in linear scale.}
  \label{fig:HNC-xcmaps}
\end{figure}

The \textit{top panel} of Figure~\ref{fig:HNC-xcmaps} 
shows all the excitation conditions for which the observed HNC $\frac{3-2}{1-0}$ and HNC $\frac{4-3}{3-2}$ 
line ratios and the intensity of the HNC $J$=3--2 line can be reproduced within $1\sigma$. 
The mean column density of HNC per line width ($N(\rm{HNC})/\Delta\upsilon~\ndv$) is shown in the
contour plot. The \textit{bottom panel} of Figure~\ref{fig:HNC-xcmaps} shows how the column density changes 
with $\nh2$ at different temperatures. The corresponding error bars ranges are 6.4\% -- 35.9\% at 30 K, 7.6\% -- 32.3\% at 50 K, 
and 8.3\% -- 26.0\% at 70 K. The error bars decrease towards the higher $\nh2$ densities.

The estimated $N(\rm{HNC})/\Delta\upsilon~\ndv$ column density per line width ranges between $10^{12.8}~\ndv$ and $10^{15.4}~\ndv$. 
The $\nh2$ densities required to reproduce these ratios range from $10^4~\3cm$ (and probably lower than that, 
when considering higher temperatures outside of our explored grid) and $10^7~\3cm$ at temperatures of about 10 K.
A summary of the corresponding optical depth $\tau$ can be found in Table~\ref{tab:xc}.

In contrast with HCN, the temperatures at which solutions can be found for HNC are limited up to about 90 K, 
for the lowest densities explored. At a density of $10^{5.5}~\3cm$, $T_K$ ranges between 10 K and 70 K.
But at densities $\ge10^6~\3cm$ only temperatures lower than 30 K are allowed.

\subsection{The physical environment of \hcop}

\begin{figure}[!tp]
  \centering
  \includegraphics[width=7cm]{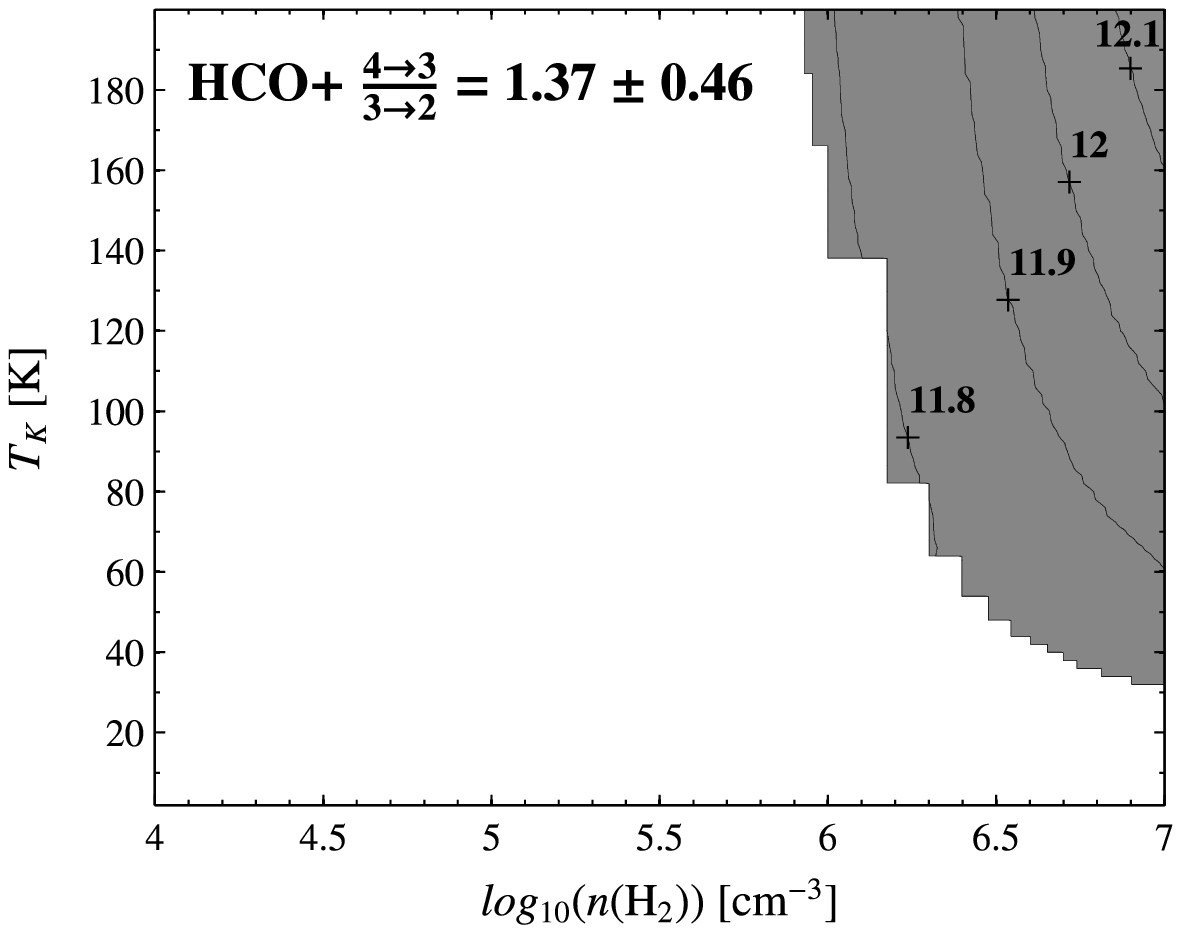}\\
  \includegraphics[width=7cm]{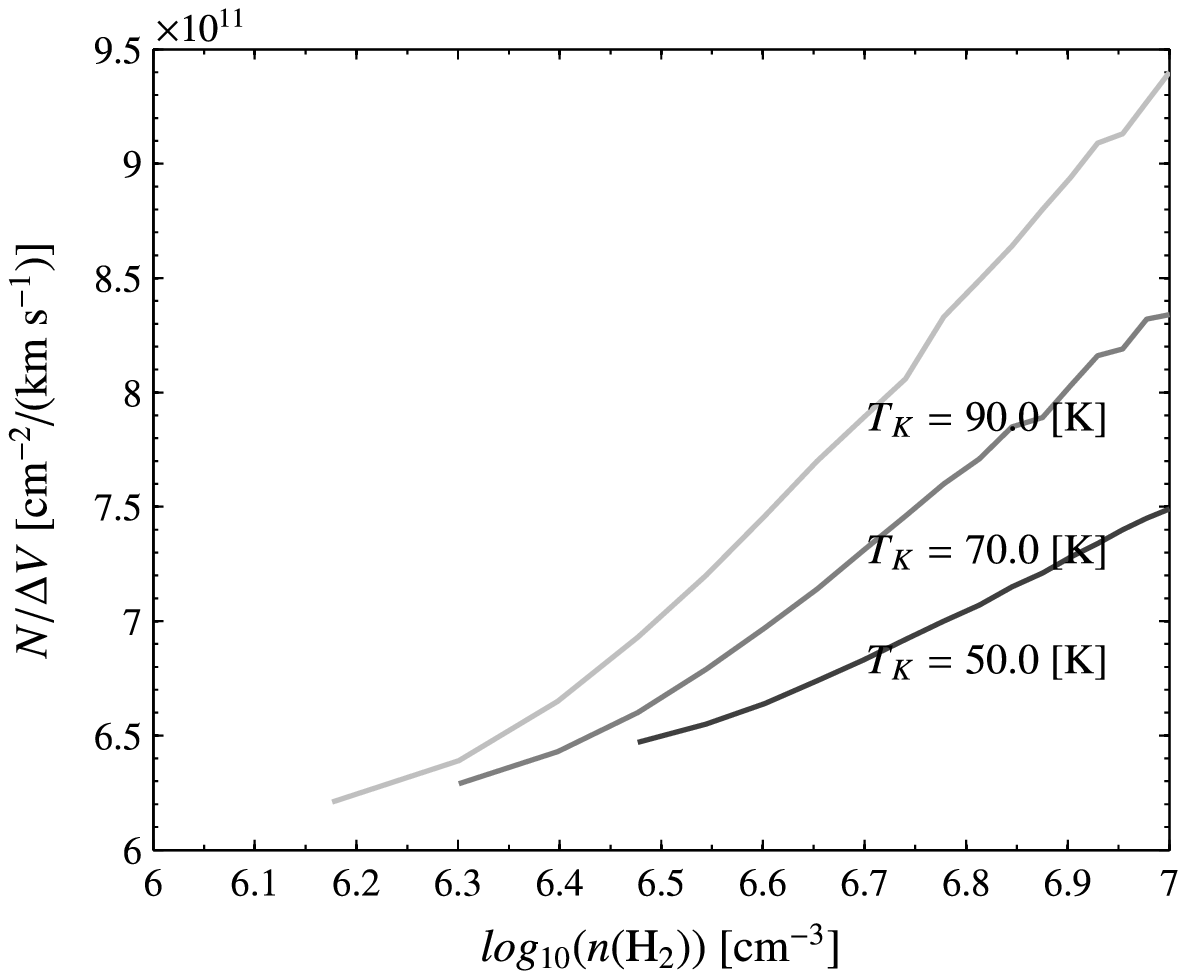}
  
  \caption{\textit{Top} - Excitation conditions modeled for the $\frac{4-3}{3-2}$ line ratio of \hcop. 
  The contour lines correspond to the mean column density of \hcop per line width ($N(\rm{HCO^+})/\Delta\upsilon~\ndv$), 
  in the region where the estimated line ratios and the $J$=3--2 line intensity reproduce,
  within $1\sigma$, the observed values.
  \textit{Bottom} - Transversal cuts at different temperatures of the column density per line width modeled above. The column densities are of the order of $10^{11}~\ndv$, in linear scale.} 
  \label{fig:HCOp-xcmaps}
\end{figure}

Using the common source size of 1.5'' and the starburst contribution factor of 0.45 in the \hcop~$J$=1--0 line, 
our model is not able to reproduce the observed \hcop~$\frac{3-2}{1-0}$ line ratio and \hcop~$J$=3--2 line intensity. 
The \textit{top panel} of Figure~\ref{fig:HCOp-xcmaps} shows the mean column density of \hcop per line width 
($N(\rm{HCO^+})/\Delta\upsilon~\ndv$) modeled for the \hcop~$\frac{4-3}{3-2}$ line ratio and \hcop~$J$=3--2 line intensity.
The \textit{bottom panel} of Figure~\ref{fig:HCOp-xcmaps} shows how the column density changes 
with $\nh2$ at different temperatures. The corresponding error bars ranges are 29.9\% -- 30.1\% at 50 K, 
and 29.3\% -- 30.0\% at 70 K, and 29.0\% -- 30.1\% at 90 K. In this case the error bars decrease towards the lower $\nh2$ densities.

The model shows that the high-$J$ lines trace gas with densities larger than 
$10^{5.9}~\3cm$ and temperatures larger than 30 K. Solutions with temperatures lower than 30 K could also
be found at densities larger than those explored in this work ($\nh2>10^7~\3cm$).
The columns range from $10^{11.9}~\ndv$ to $10^{12.2}~\ndv$, and both lines are optically thin 
($\tau\le10^{-1.2}$) over the whole range of columns (Table~\ref{tab:xc}).
The analysis of the uncertainties in the \hcop~model is discussed in the Appendix B.2.

\subsection{The physical environment of \cn}

Only the main spingroups of each transition were considered for modelling 
the physical conditions of CN. The hyperfine structure of CN is not included in the model. 
Hence, the radiative lines of CN are described just by the quantum numbers $N$ and $J$. 
The collisional data are the same as used by Fuente \etal~(\cite{fuente95}).

\begin{figure}[!tp]
  \centering
  \includegraphics[width=7cm]{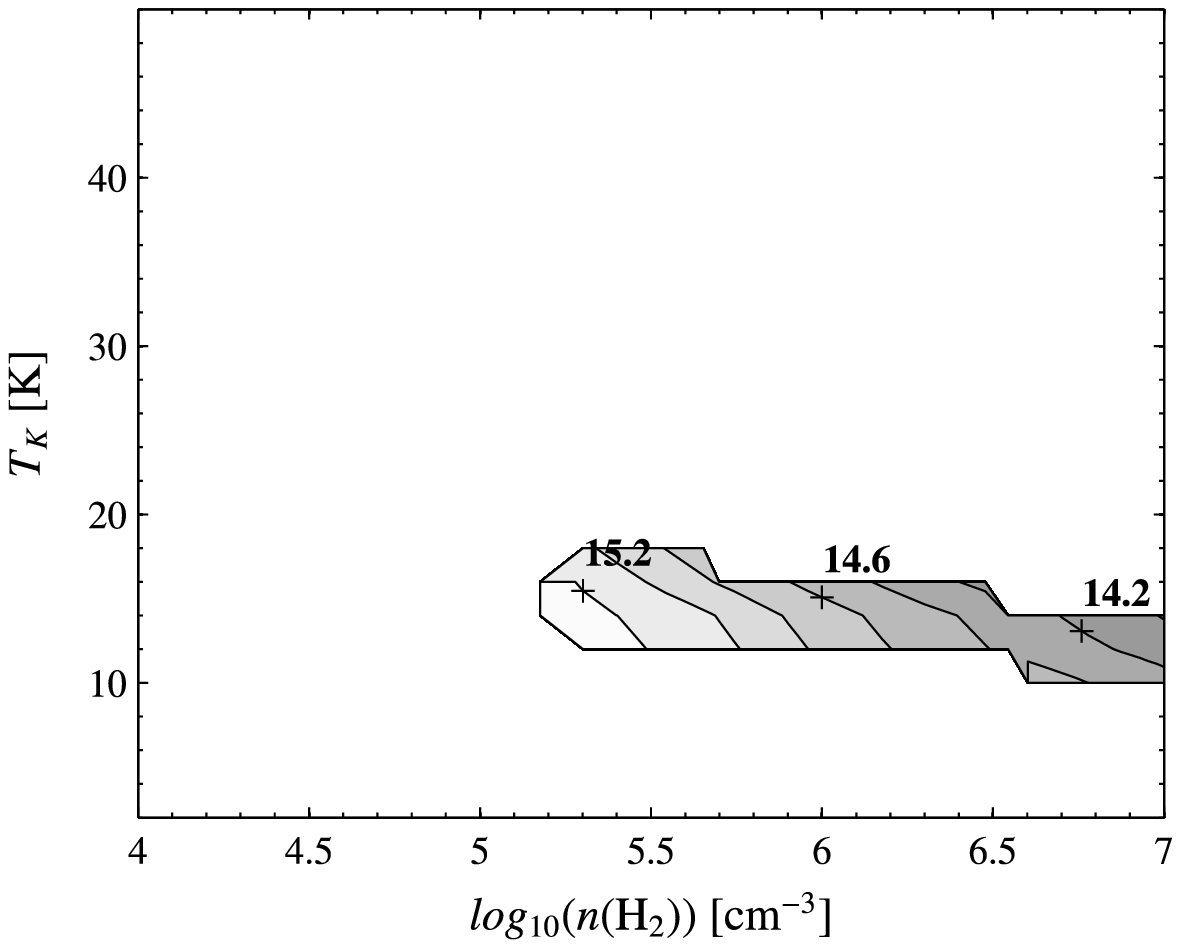}\\
  \includegraphics[width=7cm]{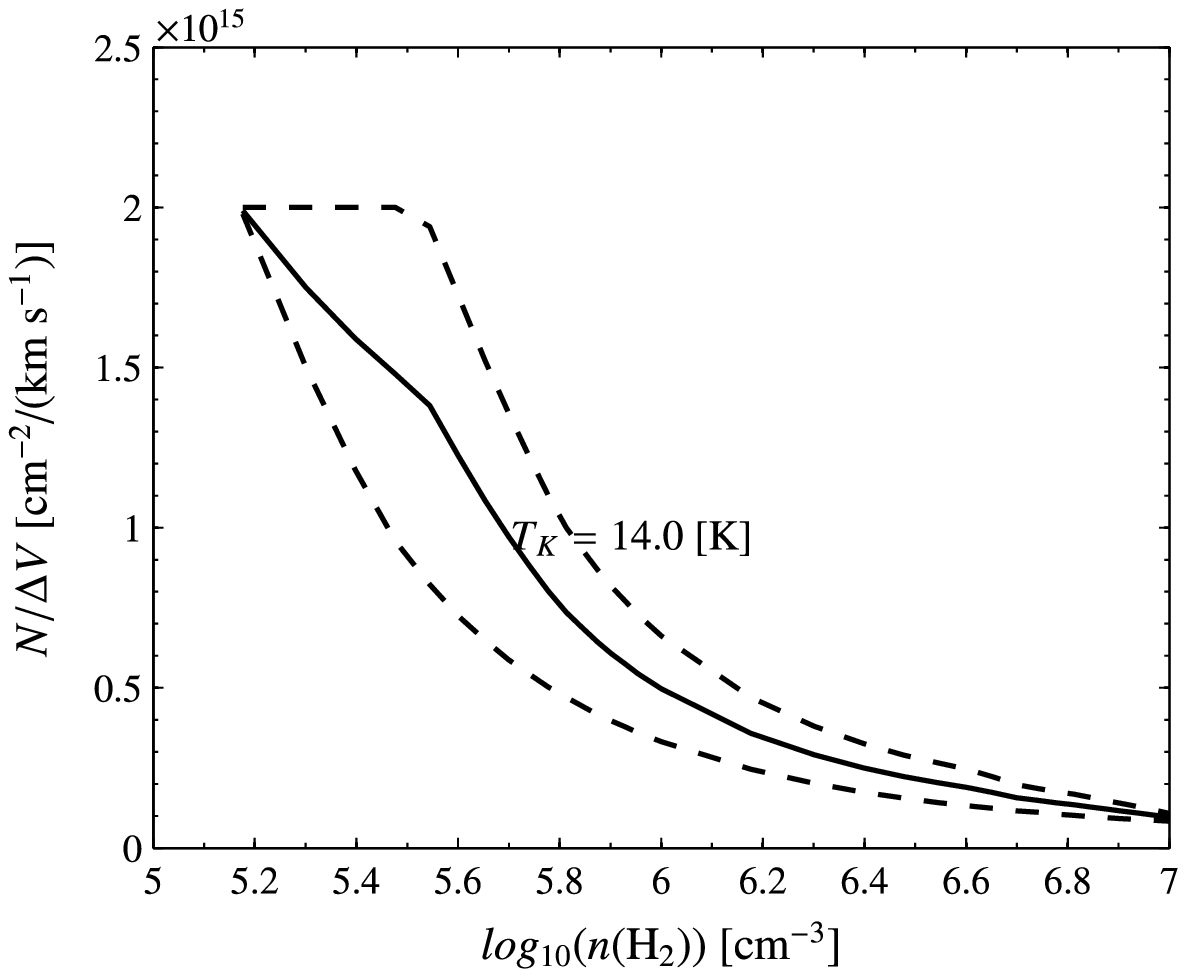}
  
  \caption{\textit{Top} - Excitation conditions modeled for the $\frac{2_{5/2}-1_{3/2}}{1_{3/2}-0_{1/2}}$ 
  and $\frac{3_{5/2}-2_{5/2}}{2_{5/2}-1_{3/2}}$ line ratios of CN.
  The contour lines correspond to the mean column density of CN per line width ($N(\rm{CN})/\Delta\upsilon~\ndv$)
  in the region where the estimated line ratios and the $N_J=2_{5/2}-1_{3/2}$ line intensity reproduce,
  within $1\sigma$, the observed values. 
  \textit{Bottom} - Transversal cut at $T_K=14$ K of the column density per line width modeled above. 
  The values are of the order of $10^{15}~\ndv$, in linear scale. The dashed lines 
  correspond to the upper and lower limits of $N(\rm{CN})/\Delta\upsilon~\ndv$.}
  \label{fig:CN-xcmaps}
\end{figure}

The excitation conditions derived from the two line ratios with the one-phase model 
(with a source size of 1.5'') are shown in the \textit{top panel} of Figure~\ref{fig:CN-xcmaps}. 
In contrast with HCN and HNC, these conditions overlap just in an small region, with a narrow 
temperature range. At the lowest density for which solutions are found ($n(\rm H_2)=10^{5.2}~\3cm$), 
the kinetic temperature of the gas $T_K$ is between 14 K and 16 K. At higher densities ($n(\rm H_2)\ge10^7~\3cm$) 
the gas can be as cold as 10 K. 

The \textit{bottom panel} of Figure~\ref{fig:CN-xcmaps} shows how the column density changes 
with $\nh2$ at $T_K=14$ K. The dashed lines correspond to the upper and lower limits 
of $N(\rm{HCO^+})/\Delta\upsilon~\ndv$. Note that for $n({\rm H_2})=10^{5.2}~\3cm$ 
and $n({\rm H_2})=10^7~\3cm$ the limits converge, which means there is only one solution at 
$T_K=14$ K for those densities.

The column densities per line width have a lower limit of about $10^{14}~\ndv$ at $n(\rm H_2)=10^7~\3cm$ 
and an upper limit of $10^{15.6}~\ndv$ at a density of $10^{4.7}~\3cm$. The corresponding optical depths
of each line can be found in Table~\ref{tab:xc}.

Besides molecular hydrogen, we also explore the effects of electrons as a second collision partner for CN. These could be significant both in PDR and XDR environments, due to enhancement of the ionization degree by radiation.
The collision rates are the same as used in Black \& Van Dishoeck (\cite{black91}), and were obtained from J.H. Black (private communication). These rates are available only for $T_K=20$ K.
 
In a PDR environment, the CN emissivity peaks at a total hydrogen column density of about $10^{21.5}-10^{22}~\2cm$. At these depths the electron abundance is $\sim10^{-5}$, and the CN column density is of the order of $10^{14}~\2cm$. We found that for densities $n(\rm{H}_2)\ge10^4~\3cm$ the effect of collisions between CN and electrons is negligible.

The only region where the electron abundance can be larger is at the edge of a PDR cloud, that is $N_{\rm H}\le10^{21.5}~\2cm$. There the electron abundance is still about four orders of magnitude lower than the total hydrogen density, but it can be at least two orders of magnitude larger than the H$_2$ density (Meijerink \& Spaans~\cite{meijerink05}). At those shallow depths, however, the column of CN is not significant ($N(\rm{CN})\le10^{10}~\2cm$). In order to boost the electron abundance to higher levels, in the region where most of the CN emission originates, we would require a gas phase carbon abundance of about $3-4$ times Solar. However, these higher abundances are not supported by other works (e.g. Kraemer \etal~\cite{kraemer98})

In an XDR, the ambient conditions along the cloud are different than those found in a PDR, with larger ($10^{-2} - 10^{-4}$) relative electron abundances (Meijerink \& Spaans~\cite{meijerink05}). In the main emitting region ($N(\rm{H})\ge10^{22}~\2cm$) the electron density $n(e^-)$ is expected to be about 10 $\3cm$ if $n(\rm{H}_2)\sim10^5~\3cm$. These densities produce changes of the order of 10\% in the excitation temperatures $T_{ex}$ of CN, with respect to those obtained when using only H$_2$ as collision partner, and the column density $N(\rm CN)$ has to decrease with about 50\% to get the same line strengths.

Hence, the effect of electrons as secondary collision partner of CN, is not important in a PDR environment. In an XDR environment, small effects can be expected.

\section{Analysis of the uncertainties}

We analyze here how the uncertainties in the input parameters of our models (4--3/3--2, 3--2/1--0 line ratios and the $J$=3--2 line intensities) propagates to the solutions we find.
The two main uncertainties in our models are the estimated common source size ($\theta_S=1.5''$) for all the molecules and transitions, and the first order estimate of the starburst contribution factor $f_{\rm SB}$ for the $J$=1--0 lines. The starburst contribution factor affects only the $\frac{3-2}{1-0}$ line ratios which in turn will modify, to some extent, the combination of temperatures ($T_K$), densities ($\nh2$) and column densities ($N$) for which solutions are found. On the other hand, the source size affects mostly the $J$=3--2 line intensities (used to constrain the radiative transfer models) which in turn affect mostly the range of temperature and column densities of the solutions. The source size is also present in the 4--3/3--2 and 3--2/1--0 line ratios, as we correct these lines for beam dilution. However, because the source size is about one order of magnitude smaller than the size of the respective beams, its effect in the line ratios is negligible.

The solutions for the HCN and HNC line ratios are less sensitive to these two parameters since the range of temperatures and densities for which solutions exist is large enough (which allows for more flexibility in the $T_K$ vs $\nh2$ space) and because the $\frac{4-3}{3-2}$ line ratios, which basically define and constrain the overlap with the solutions found for the ratio between the lower-$J$ lines, are independent of $f_{\rm SB}$. The effect is reflected mostly in the column densities per line width due to changes in the source size. 

\begin{figure}[!t]
  \centering
  \includegraphics[width=7cm]{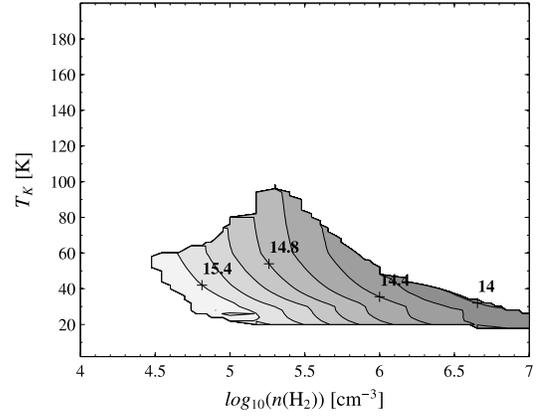}
  
  \caption{Excitation conditions modeled for the $\frac{2_{5/2}-1_{3/2}}{1_{3/2}-0_{1/2}}$ 
  and $\frac{3_{5/2}-2_{5/2}}{2_{5/2}-1_{3/2}}$ line ratios of CN, using a source size of 1''.
  The contour lines are as defined before. Considering this smaller source size, the range of temperatures 
  where solutions are found can go up to 100 K, and densities can be as low as $10^{4.5}~\3cm$. }
  \label{fig:CN-xcmaps-new}
\end{figure}

\subsection{Uncertainties in \cn}

The solutions for CN are particularly sensitive to the source size. As mentioned before, the starburst contribution factor affects only the ratio between the lower $J$-lines. The range of $T_K$ and $\nh2$ for which we find solutions for the CN $\frac{3_{5/2}-2_{5/2}}{2_{5/2}-1_{3/2}}$ line ratio is smaller than that for the $\frac{2_{5/2}-1_{3/2}}{1_{3/2}-0_{1/2}}$ ratio. Hence, the final solutions for CN (given by the overlap between the solutions found separately for the ratios between the low-$J$ lines and the high-$J$ lines) is constrained by the solutions found for the $\frac{3_{5/2}-2_{5/2}}{2_{5/2}-1_{3/2}}$ line ratio. 

Because the solutions found for the CN are already constrained, little changes in the source size have a larger impact than for HCN and HNC. If we assume a larger source size of 2'', the solutions for CN are restricted to an small range of temperature around $10\pm4$ K and for densities between $10^{5.7}~\3cm$ and $10^{6.6}~\3cm$. On the other hand, if we assume an smaller source size of 1'', the solutions for CN line ratios can be found in a larger region of $T_K$ and $\nh2$ than those found for a source size of 1.5'', as shown in Figure~\ref{fig:CN-xcmaps-new}. 
The smaller source size of 1'' increases the estimated CN $N_J=2_{5/2}-1_{3/2}$ radiation temperature by a factor 2.25, which is reflected mostly in the range of kinetic temperatures at which we can find solutions for the observed line ratios. When using $\theta_S=1.5''$ the maximum temperature where solutions can be found is 20 K, at a density of $10^{5.3}~\3cm$ (Figure~\ref{fig:CN-xcmaps}). But when using an smaller source size (Figure~\ref{fig:CN-xcmaps-new}) the temperature range can go \textit{from} 20 K up to 100 K, at the same density.

\begin{figure}[!t]
  \centering
  \includegraphics[width=7cm]{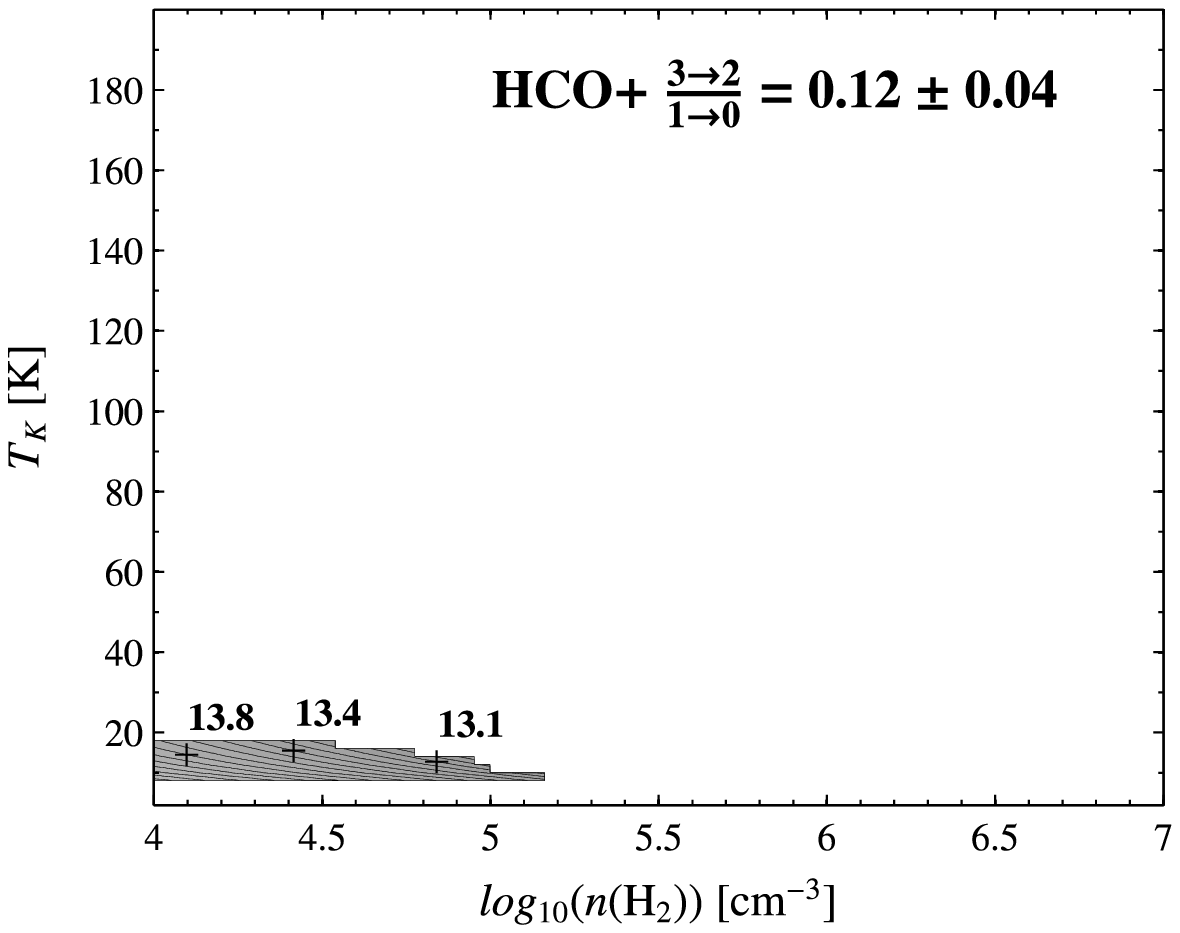}\\
  \includegraphics[width=7cm]{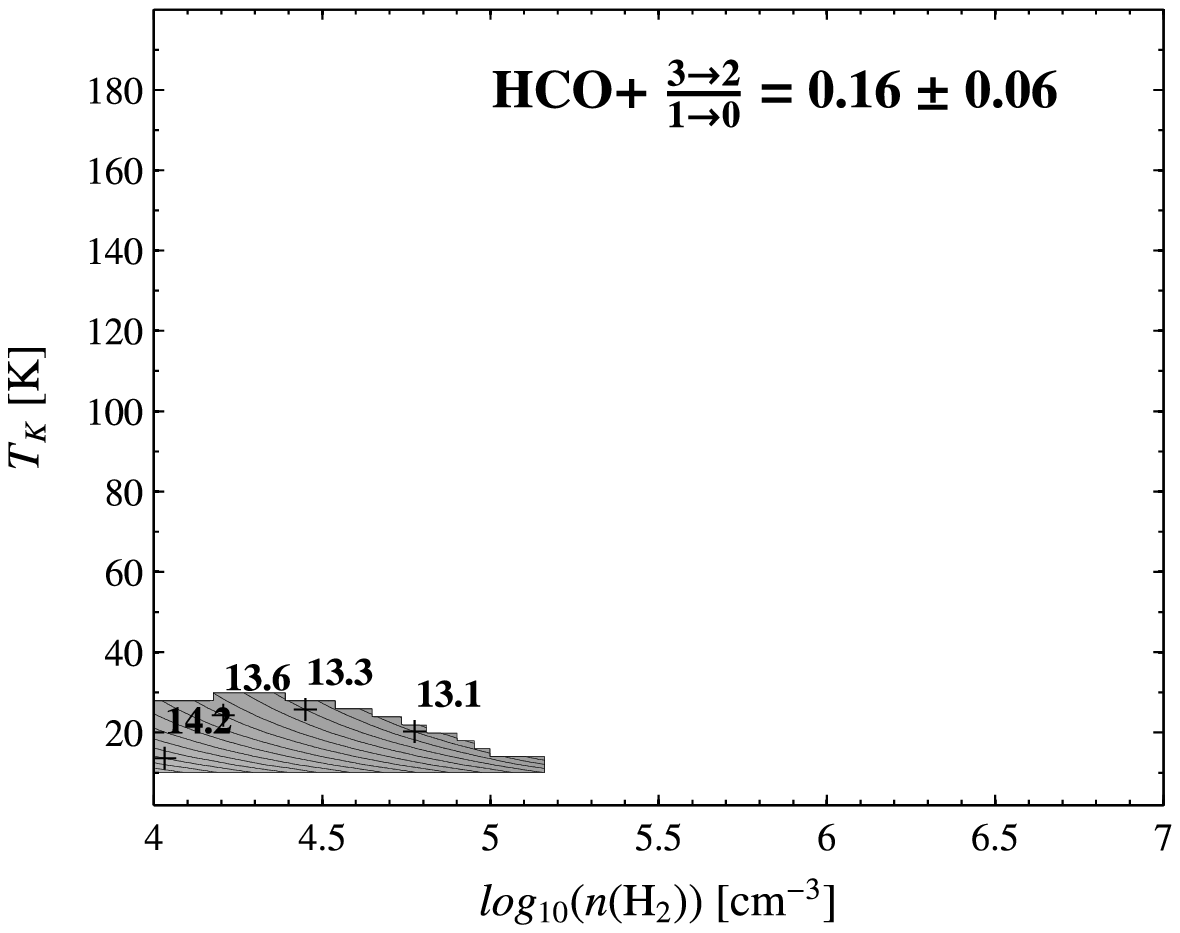}\\
\includegraphics[width=7cm]{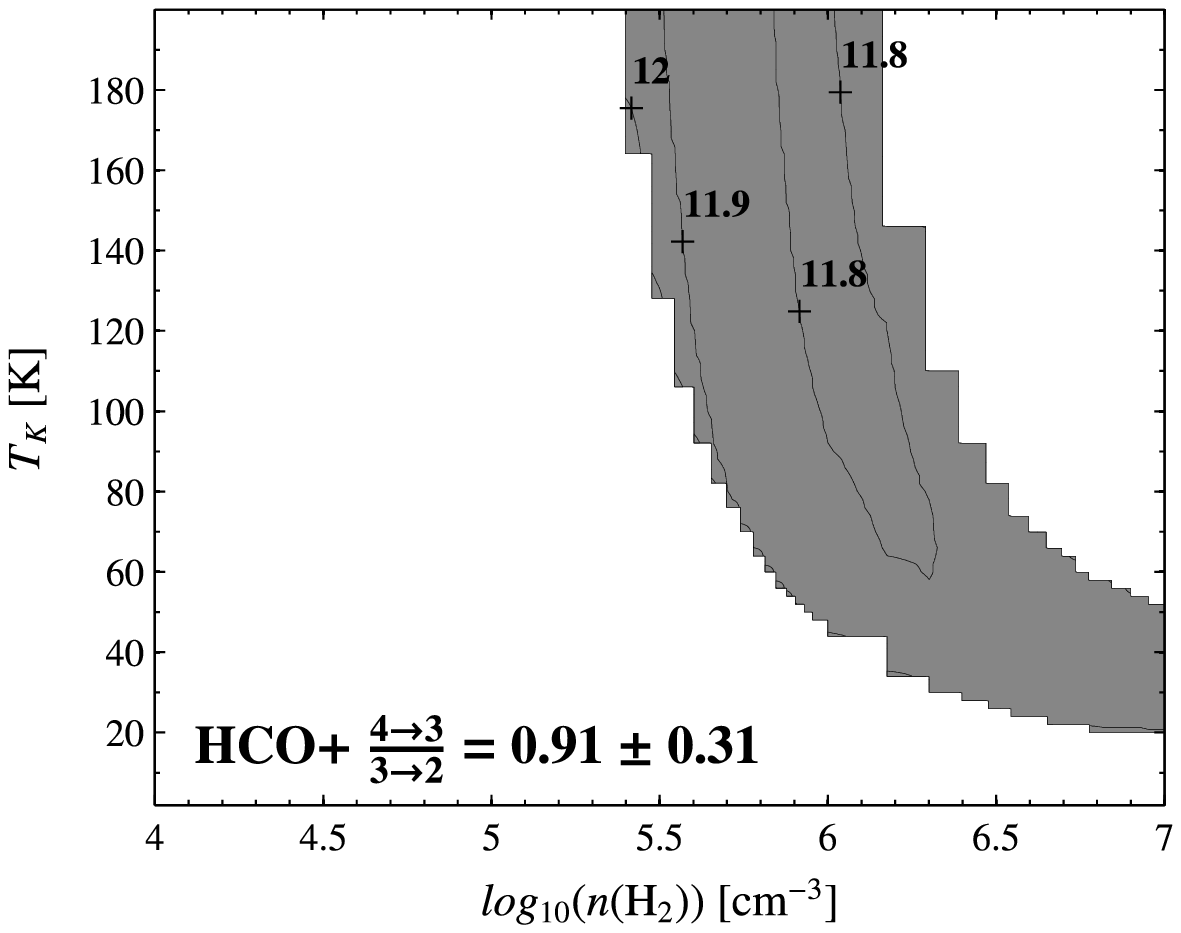}\\
  
  \caption{\textit{Top} - Excitation conditions modeled for the $\frac{3-2}{1-0}$ line ratio of \hcop, when considering the estimated starburst contribution factor $f_{SB}=0.45$, but a larger source size $\theta_{S}=2''$. \textit{Middle} - Excitation conditions modeled for the $\frac{3-2}{1-0}$ line ratio of \hcop, when considering a larger starburst contribution factor $f_{SB}=0.60$, and the common source size $\theta_{S}=1.5''$. \textit{Bottom} - Excitation conditions modeled for the $\frac{4-3}{3-2}$ line ratio of \hcop, when reducing the \hcop~$J$=4--3 line intensity by a factor 1.5. The higher transition $J$=4--3 still traces mostly the dense and warm gas ($\nh2\ge 10^{5.4}~\3cm$, $T_K\ge20$ K), whereas the $\frac{3-2}{1-0}$ line ratio indicates that those lines would trace less dense ($\nh2<10^{5.2}~\3cm$) and cold (10 K $< T_K <$ 30 K) gas.} 
  \label{fig:HCOp-xcmaps-uncertain}
\end{figure}

\subsection{Uncertainties in \hcop}

On the other hand, the \hcop~$\frac{3-2}{1-0}$ is affected by both, $\theta_S$ and $f_{\rm SB}$, and the fact that we do not find solutions for this ratio may be due to the uncertainties in these two parameters. We explored different alternatives and we found that our model can reproduce the observed ratio, and \hcop~$J$=3--2 line intensity, if we assume either a larger source size ($\theta_S=2''$), or a larger starburst contribution factor ($f_{\rm SB}=0.60$). The results are shown in the \textit{top} and \textit{middle} panels of Figure~\ref{fig:HCOp-xcmaps-uncertain}. 
A larger starburst contribution factor increases the \hcop~$\frac{3-2}{1-0}$ line ratio, and allows for solutions at slightly higher temperatures than assuming a larger source size. Note that in both cases solutions for densities $<10^4~\3cm$ are also possible. This result, however, rises a new question since these solutions do not overlap with the solutions found for the \hcop~$\frac{4-3}{3-2}$ ratio (Figure~\ref{fig:HCOp-xcmaps}).
 
The fact that the ratio between the higher transition lines traces denser ($\nh2>10^{5.9}~\3cm$) and warmer ($T_K>30$ K) gas than what the lower transition lines would indicate, and that they do not overlap either with the solutions found for HCN, HNC or CN, may raise some skepticism about the \hcop~$J$=4--3 line. So we also explored the results of our model, assuming that the \hcop~$J$=4--3 line could be somehow overestimated (e.g., undetected calibration problems, or intrinsic instrumental differences, like sensitivities, between the JCMT and the IRAM 30m telescopes). 

In section 2 we mentioned that HARP observations of the calibration source showed less flux than the reference spectra observed with the former receiver B3. Those differences account for factors between 1.1 and 1.5, that where used to correct each of our scans before adding them up. So, if we reduce the final \hcop~$J$=4--3 line by a factor 1.5 (which would be the worst case scenario) the \hcop~$\frac{4-3}{3-2}$ line ratio reduces to $0.91\pm0.31$ and the new solutions would be as shown in the \textit{bottom} panel of Figure~\ref{fig:HCOp-xcmaps-uncertain}. It can be seen that, even in this worst case scenario, there would not be an overlap between the solutions found for the ratio between the lower $J$ lines and those found for the higher $J$ lines.

This result would imply that the lower and higher $J$-lines of \hcop~trace different gas phases, and a single-phase model may not be the most appropriate to reproduce the observed \hcop~ratios and intensities. Hence, a two-phase model, where different sizes of the emitting region could be seen by different $J$ lines, may be a better approach for \hcop. In fact, Krips et al. (2008) mentioned that the size of the emitting region decreases with increasing $J$ line, according to recent SMA data. A multiple-phase model can also be applied to the HCN, HNC and CN molecules. Among molecules, those which are easily dissociated should have small sizes. Hence, a decreasing source size with increasing $J$ line seems natural.
However, the only way to constrain these models would be by using interferometer maps (fluxes, beam deconvolution, etc.) to accurately estimate the actual source size seen by the different $J$ lines. Nevertheless, this is a task that is beyond the scope of this work.

\section{Line intensity and abundance ratios}

The line intensity ratios, with respect to HCN, are summarized in Table~\ref{tab:molecule-ratios}.
The intensities used correspond to the peak antenna temperature of the main component of the gaussian fits, corrected for starburst contribution, beam efficiency, and beam dilution, assuming a source size of 1.5'' for all the lines.
We also find that most of the line intensity ratios (the \hcop/\hcn $J$=4--3 ratio is the exception that is discussed below) decrease with increasing rotational quantum number $J$, similar to what was found by Krips \etal\ (2008) for the \hcn/CO ratio. This was also noticed for the \hnc/\hcn\ $J$=1--0 and $J$=3--2 line ratios observed in NGC~1068, and other Seyfert galaxies, by P\'erez-Beaupuits \etal\ (2007). Since the beam sizes are comparable, this could be an indication that the higher-$J$ levels of CN and HNC are less populated than those of HCN.

This is what we would expect assuming collisional excitation of the molecules, where $T_{ex}$ is proportional to $\tau\times n(\rm H_2)$. From the excitation conditions modeled in Section \S4.3, the HNC column densities, and hence $\tau_{\rm HNC}$, tend to be lower than that of HCN. On the other hand, even if the column densities that we find for CN are as high (or higher in some cases) than the columns found for HCN, the optical depth $\tau_{\rm CN}$ tends to be lower than $\tau_{\rm HCN}$ because of the higher (fine and hyperfine) splitting of the rotational levels of CN. Thus, the high energy levels of CN and HNC would be less populated than those of HCN.

In the following sections we address the relative abundance issue by estimating the abundance ratio as the ratio between the column densities estimated from the radiative transfer models described before.

\begin{table}[!t]
      \caption[]{Line intensity ratios between molecules.}
         \label{tab:molecule-ratios}
	\centering
         \begin{tabular}{lccc}
            \hline\hline
            \noalign{\smallskip}
            Molecules &  & Transition$^{\mathrm{a}}$ & \\
                      & $J$=1--0 & $J$=3--2 & $J$=4--3 \\ 
            \noalign{\smallskip}
            \hline
            \noalign{\smallskip}
	    $\frac{\rm HNC}{\rm HCN}$ & 0.71$\pm$0.45 & 0.26$\pm$0.08 & 0.16$\pm$0.05 \\
            \noalign{\smallskip}
            $\frac{\rm CN}{\rm HCN}$ & 0.89$\pm$0.69 & 0.54$\pm$0.15 & 0.14$\pm$0.04 \\ 
            \noalign{\smallskip}
            $\frac{\rm HCO^+}{\rm HCN}$ & 1.10$\pm$0.59 & 0.13$\pm$0.04 & 0.36$\pm$0.10 \\ 
            \noalign{\smallskip}	    	    
            \hline
         \end{tabular}
\begin{list}{}{}
\item[${\mathrm{a}}$)] The lines correspond to $J$=1--0, $J$=3--2 and $J$=4--3, 
for HCN, HNC and \hcop. In the case of CN, the lines are actually 
$N_J=1_{3/2}-0_{1/2}$, $N_J=2_{5/2}-1_{3/2}$ and $N_J=3_{5/2}-2_{5/2}$, respectively.
\end{list}
\end{table}

\subsection{\hnc/\hcn}

\begin{figure}[!t]
  \centering
  \includegraphics[width=7cm]{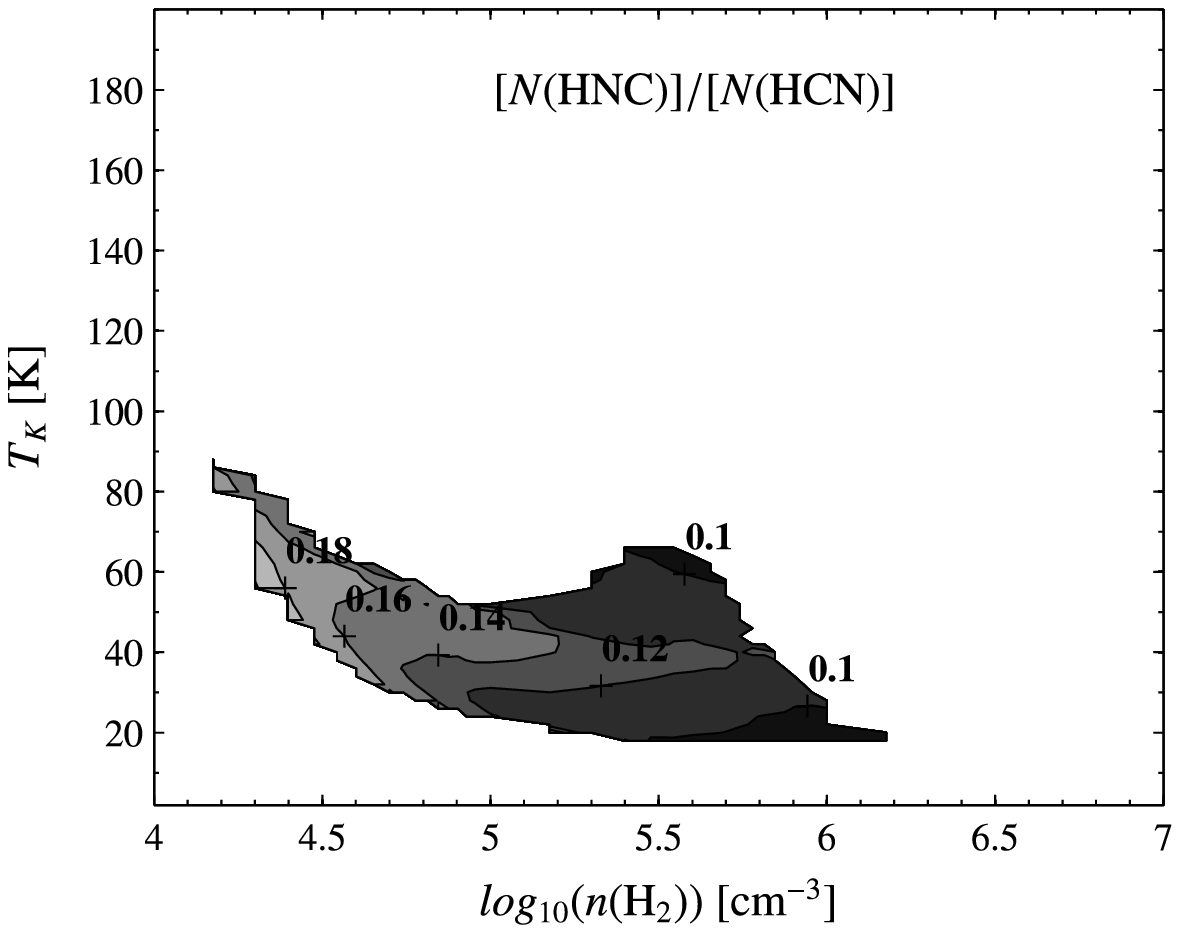}\\
  \includegraphics[width=7cm]{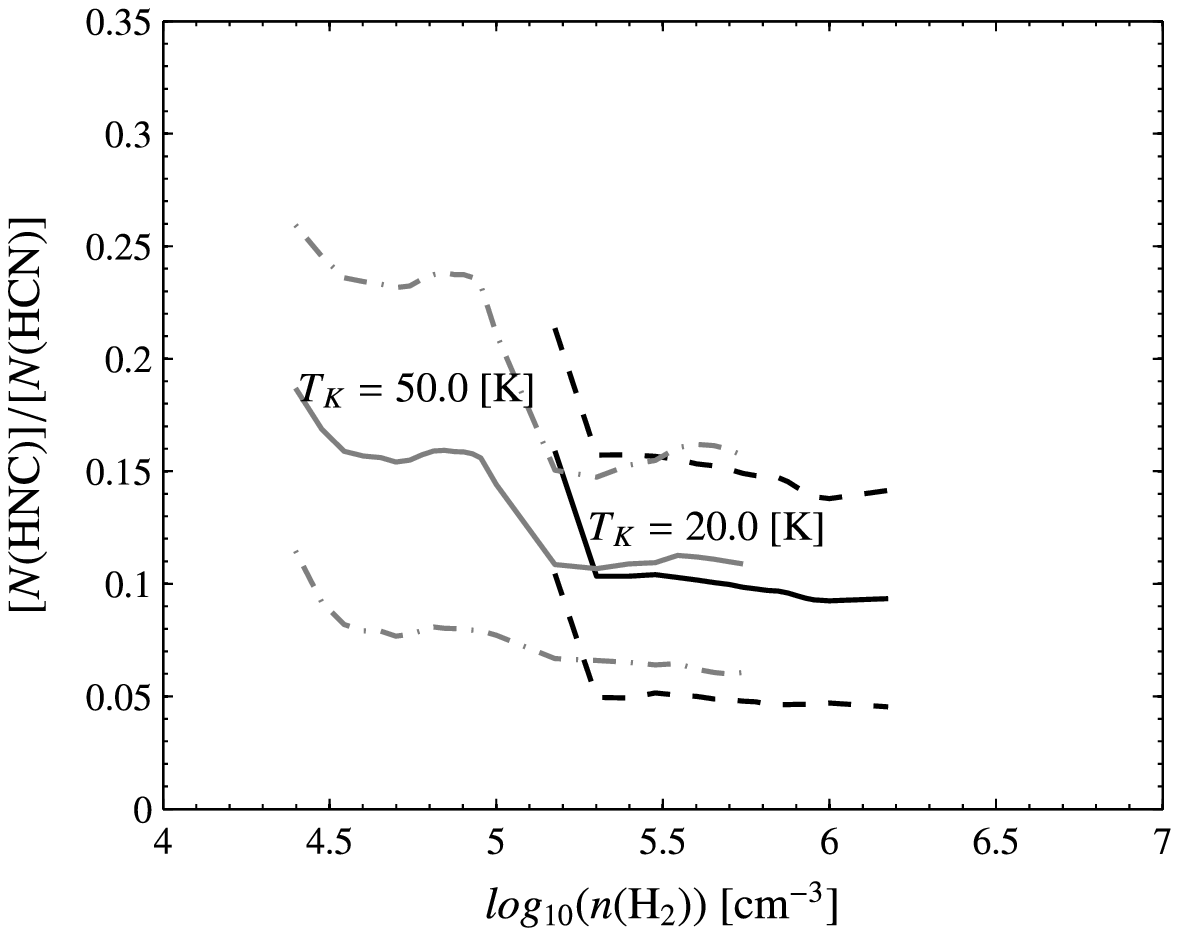}
  
  \caption{\textit{Top} - Ratios between the mean column densities of HNC and HCN ($N(\rm{HNC})/N(\rm{HCN})$).
  \textit{Bottom} - Transversal cut of the ratios at different temperatures. The \textit{dashed} and \textit{dashed-dot} lines 
  correspond to the upper and lower limits of the ratios at 20 K and 50 K, respectively.}
  \label{fig:NHNC-NHCN-ratios}
\end{figure}

The range of temperatures where we can analyze the \hnc/\hcn~ratio is limited by the solutions found for HNC, which go up to $\sim90$ K at the lowest density explored ($\nh2=10^4~\3cm$). The \textit{top panel} in Figure~\ref{fig:NHNC-NHCN-ratios} shows the mean $N({\rm HNC})/N({\rm HCN})$ column density ratio for the temperatures and densities where the solutions found for each molecule (top panels of Figs.~\ref{fig:HNC-xcmaps} \&~\ref{fig:HCN-xcmaps}) overlap. The mean column density ratio ranges between 0.10 and 0.18, with errors that vary between 30\% and 55\% of the mean value. Note that similar ratios can be found at two different temperatures for a particular density. The \textit{bottom panel} in Figure~\ref{fig:NHNC-NHCN-ratios} shows the mean value, and corresponding upper and lower limits, of the $N({\rm HNC})/N({\rm HCN})$ ratio at 20 K and 50 K. At these temperatures the ratios are quite similar (within 15\%) for the density range $\nh2=10^{5.3-5.7}~\3cm$.

The fact that the line intensity ratios (Table~\ref{tab:molecule-ratios}) are also lower than unity, indicates that the bulk of the HNC and HCN emission emerges from warm gas ($T_K>30$ K). This agrees with observations in the vicinity of the hot core of Orion KL, and experimental and theoretical data, where the HNC/HCN line ratio decreases as the temperature and density increase (e.g. Schilke \etal\ 1992; Talbi \etal\ 1996; Tachikawa \etal\ 2003).

On the other hand, the ratios $N(\hnc)/\textit{N}(\hcn)<1$ estimated with our models cannot be directly interpreted as a signature of a pure PDR or XDR environment in the CND of NGC~1068. The HNC abundance can be decreased due to temperatures higher than traditionally expected, produced deep inside a molecular cloud by mechanisms other than radiation, like turbulence and shocks (Loenen \etal~\cite{loenen08}). 
If the temperature is higher than 100 K, the conversion of HNC into HCN is more efficient and HNC is suppressed (Schilke \etal~\cite{schilke92}, Talbi \etal~\cite{talbi96}). These high temperatures are not found in regions where the abundance of HCN and HNC is high enough to be detected, for traditional PDR or XDR models (Meijerink \& Spaans \cite{meijerink05}). 
The high SiO abundance observed in the CND of NGC~1068 can be a direct evidence of the possible contributions from mechanical heating (shocks) and dust grain chemistry induced by X-rays (Garc\'ia-Burillo \etal\ 2008). Hence, a more elaborated PDR/XDR model that includes both mechanical heating and grain surface chemistry will be needed to further understand the results of our radiative transfer models.

\subsection{\cn/\hcn}

\begin{figure}[!t]
  \centering
  \includegraphics[width=7cm]{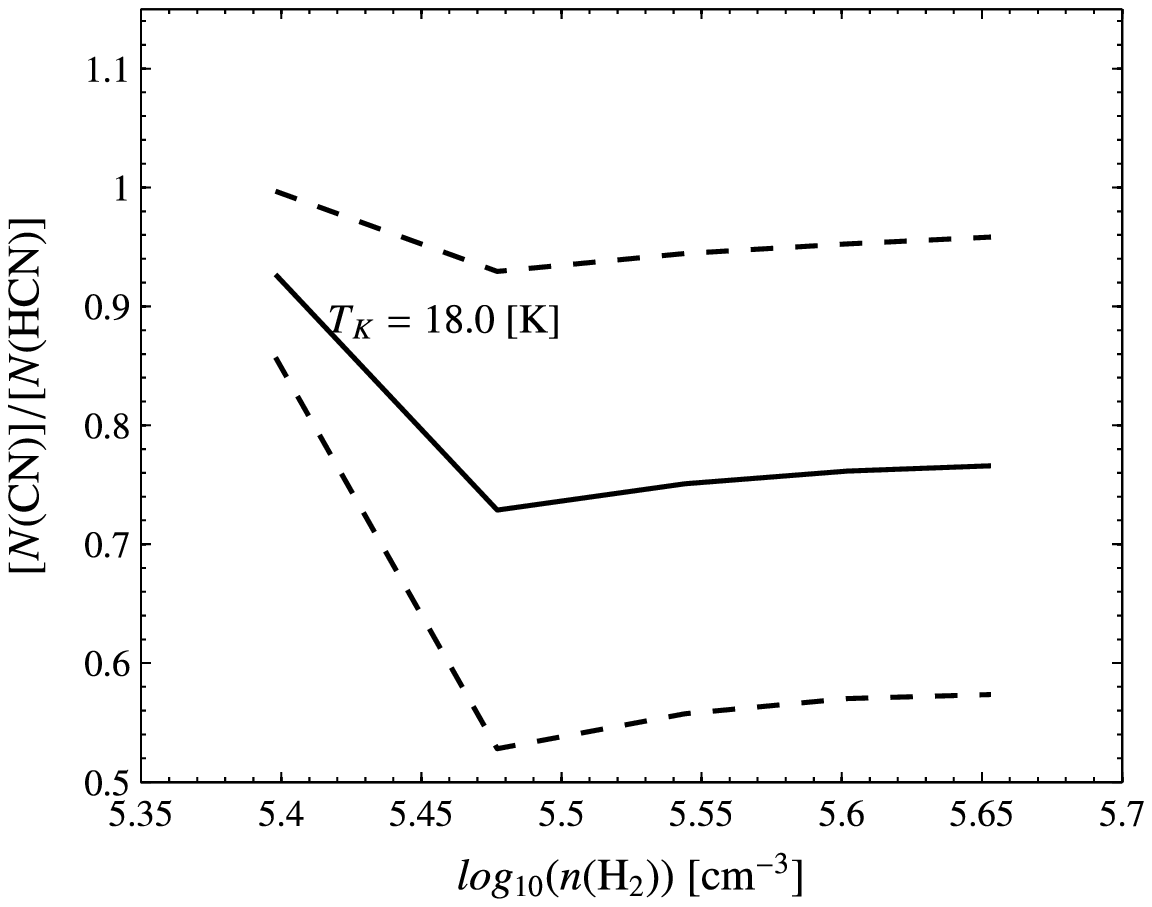}\\
  \includegraphics[width=7cm]{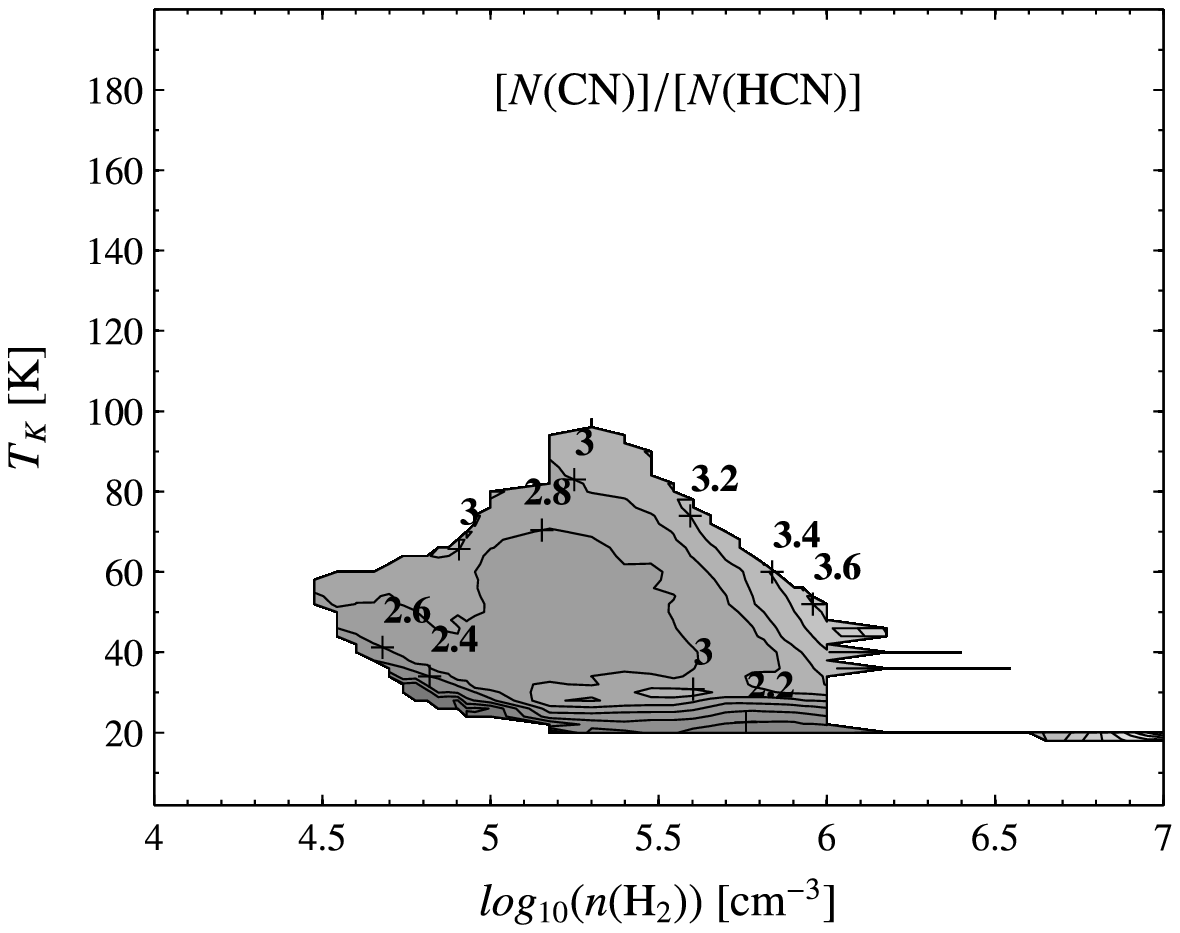}\\
  \includegraphics[width=7cm]{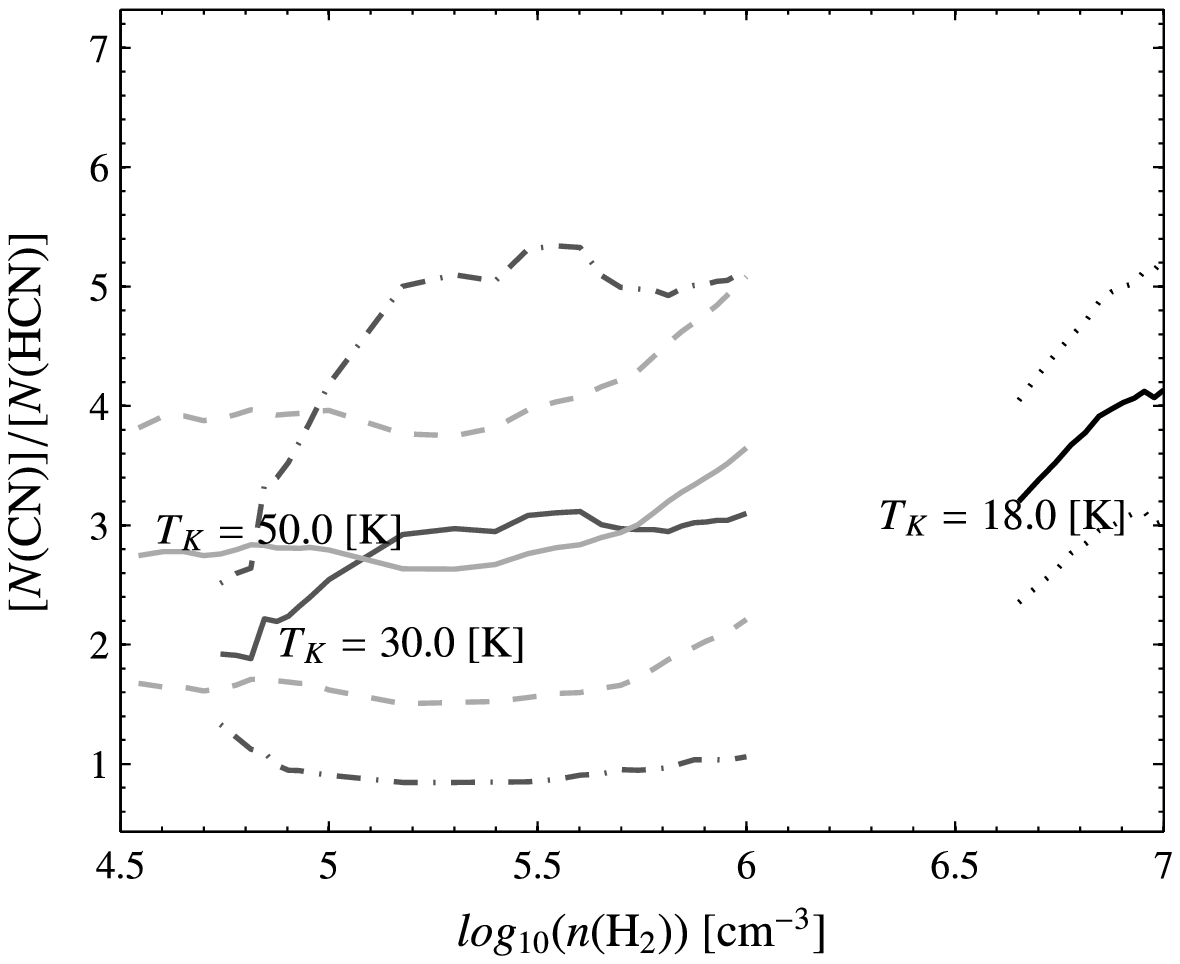}
  
  \caption{\textit{Top} - The $N({\rm CN})/N({\rm HCN})$) column density ratios for the small range of density at $T_K=18$ K, where the physical conditions found for CN and HCN overlap when using the one-phase model with $\theta_S=1.5''$. 
  \textit{Middle} - Ratios between the mean column densities of CN and HCN, when using a source size $\theta_S=1''$ for CN. \textit{Bottom} - Transversal cuts, at different temperatures, of the ratios showed in the middle panel. The \textit{dotted}, \textit{dashed-dotted} and \textit{dashed} lines correspond to the upper and lower limits of the ratios at 18 K, 30 K and 50 K, respectively.}
  \label{fig:NCN-NHCN-ratios}
\end{figure}

When using the one-phase model with a source size $\theta_{S}=1.5''$, the physical conditions estimated for the CN and HCN molecules (top panel in Figs.~\ref{fig:CN-xcmaps} \&~\ref{fig:HCN-xcmaps}) overlap \textit{only} in an small region around $\nh2\sim10^{5.5}~\3cm$ and $T_K\sim20$ K. The \textit{top panel} in Figure~\ref{fig:NCN-NHCN-ratios} shows the mean $N({\rm CN})/N({\rm HCN})$ column density ratio and the corresponding upper and lower limits. The maximum ratio of $0.93\pm0.07$ is found at $\nh2\sim10^{5.4}~\3cm$. The ratio decreases almost linearly with density (while the uncertainty increases), and it reaches the minimum ratio of $0.73\pm0.2$, with a larger uncertainty of $\sim29\%$, at a density of $\nh2\sim10^{5.5}~\3cm$. The ratio then increases slowly, with a constant uncertainty, up to $0.76\pm0.2$ at $\nh2\sim10^{5.7}~\3cm$.

If we assume a larger source size $\theta_{S}=2''$ for CN, the solutions obtained do not overlap with those found for HCN. However, when using an smaller source size $\theta_{S}=1''$ for CN, there is a large overlap in the physical conditions found for these molecules (top panel in Figs.~\ref{fig:CN-xcmaps-new} \&~\ref{fig:HCN-xcmaps}). The \textit{middle panel} in Figure~\ref{fig:NCN-NHCN-ratios} shows the mean $N({\rm CN})/N({\rm HCN})$ column density ratio for the new model. The new mean column density ratios now range between $\sim2$ and $\sim4$, with errors that vary between 20\% and 50\% of the mean value. The maximum mean ratios ($\ge3.6$) are found at a density of $\sim10^6~\3cm$ and temperatures between 40 K and 60 K, while the minimum mean ratios ($<2.3$) can be found at a larger density range of $\nh2\sim10^{5-6}~\3cm$ but at lower temperatures ($T_K<30$ K).

The \textit{bottom panel} in Figure~\ref{fig:NCN-NHCN-ratios} shows the mean value, and corresponding upper and lower limits, of the $N({\rm CN})/N({\rm HCN})$ ratio at 18 K, 30 K and 50 K. At $T_K=18$ K the new ratios (with uncertainties ranging from 23.6\% to 26.5\%) are larger (ranging between 3 and 4) and are found at a higher density range ($\nh2>10^{6.6}~\3cm$) than in the 1.5'' source size model. On the other hand, at densities between $10^5~\3cm$ and $10^6~\3cm$ the ratios found for 30 K and 50 K are very similar (within 10\%), but the uncertainties of the ratios at $T_K=30$ K are larger (63.6\%--72.4\%) than those for $T_K=50$ K (39.0\%--43.7\%). These larger uncertainties for the ratios at $T_K=30$ K imply that the corresponding upper limits increase to values $>5$ at a density $\nh2\sim10^{5.6}~\3cm$. Note that the uncertainties at $T_K=30 K$, and hence the upper limits, decrease for densities $<10^{5.1}~\3cm$. 

The mean column density ratios estimated with $\theta_{S}=1.5''$ can be easily found in an XDR environment, but the predominance of this component cannot be concluded from the $N({\rm CN})/N({\rm HCN})$ ratio only, since ratios $\sim1.0$ are also expected in a PDR component (Lepp \& Dalgarno 1996, Meijerink \etal~\cite{meijerink07}). The mean column density ratios $2\lesssim N({\rm CN})/N({\rm HCN})\lesssim 4$ estimated with $\theta_{S}=1''$, are tentatively more consistent with an XDR/AGN environments (Lepp \& Dalgarno 1996, Meijerink \etal~\cite{meijerink07}). However, these and higher [CN]/[HCN] abundance ratios have also been found in PDR/starburst scenarios (e.g. Fuente \etal\ 2005). The fact that we find a relatively low $N({\rm CN})/N({\rm HCN})$ column density ratio, with respect to what would be expected in a pure XDR scenario, could be explained by an overabundance of \hcn\ due to the grain-surface chemistry suggested by Garc\'ia-Burillo \etal\ (2008).

Nevertheless, assuming an smaller emitting region for CN than for HCN introduces a new question regarding the chemistry/physics driving the formation (or destruction) of these two molecules. If this is the case, then we would need to explain why CN is absent in the hypothetically more extended region covered by HCN. Perhaps this scenario could also be explained by the possible contributions from mechanical heating and dust grain chemistry suggested in Loenen \etal\ (2008) and Garc\'ia-Burillo \etal\ (2008). However, exploring this alternative would require high resolution maps of at least SiO, HCN, CN and HNC, in addition to a composite mechanical heating, X-rays and dust-grain chemistry model, to properly account for the different contributing scenarios. This is an study that can be addressed in a follow up work.

\subsection{\hcop/\hcn}

Figure~\ref{fig:HCOp-xcmaps} indicates that the emission from the high-$J$ \hcop~lines emerge from gas that does not co-exist with HCN, HNC and CN, in the nuclear region of NGC~1068.
The column density ($10^{11.9}-10^{12.2}~\ndv$) estimated from the \hcop$\frac{4-3}{3-2}$ ratio, and the possible solutions found for the \hcop$\frac{3-2}{1-0}$ ratio (top panels of Figure~\ref{fig:HCOp-xcmaps-uncertain}), also indicates that the warmer and denser gas traced by the high-$J$ lines is only an small fraction (0.5\% -- 10\%) of the total \hcop~gas. Most of it is confined to the lower transitions.

The main reason for the lack of co-existence is the \hcop\ $J$=4--3 line. The possible solutions found for the \hcop$\frac{3-2}{1-0}$ ratio, considering the uncertainties, would be consistent, in terms of density, with the \hcn~and \hnc~molecules, albeit at somewhat lower temperature. However, if we consider a larger source size of about 2'' (as shown in Section 4.3), the \hcop~$J$=3--2 line intensity will decrease, and solutions for temperatures up to 30 K (at densities of a few times $10^4~\3cm$) will be possible, and the solutions for the high-$J$ ratio will require just slightly lower ($N(\rm HCO^+)\sim10^{11.7-12.0}~\ndv$) column densities per line width. Hence, the \hcop~$J$=4--3 line seems to indicate a different gas phase. 

Krips \etal~(\cite{krips08}) found that the HCN/CO line intensity ratios decrease with increasing rotational quantum number $J$, for AGN dominated galaxies, including NGC~1068. We find the same trend in the HNC/HCN and CN/HCN line intensity ratios (Table \ref{tab:molecule-ratios}). The \hcop/HCN ratio, however, defies this trend. Interestingly, the higher-$J$ levels of \hcop\ may be more populated than its lower levels due to a local X-ray source (Meijerink \etal~2007). This result is consistent with an XDR, given that in strongly irradiated dense XDRs the \hcop~column builds up with depth to high values before the HCN does. Hence, the column weighted temperature of the \hcop\ molecule is higher, from which the \hcop~$J$=4--3 line benefits. The HCN behavior with depth is more gradual, avoiding the strong separation between the low and high-$J$ lines (e.g., figure 9 of Meijerink \& Spaans 2005).

The $J$=4--3 line ratio between the peak intensities of HCN and \hcop\ (the inverse value is shown in Table 6) is about $\sim2.7$, and is consistent with the ratio reported by Kohno \etal\ (2001). Instead, the velocity-integrated intensity ratio $\frac{I(HCN)}{I(\rm HCO^+)} J=4-3$ is $\sim3.7$ (from Table 3), which is larger than the ratio between the peak intensities due to the smaller line width of the \hcop\ $J$=4--3 line. This number is right above the maximum value shown in Fig.3 of Kohno (2005). Interestingly, this places NGC~1068 in a distinguished position within the Kohno diagram, among their pure AGNs.

\end{appendix}

\end{document}